\documentclass[12pt]{iopart}
\pdfoutput=1
\usepackage{graphicx}
\usepackage{placeins}
\usepackage{amssymb}
\usepackage{grffile}

\newcommand{\dd}{{\rm d}}

\newcommand{\one}{\underline{1}}

\newcommand{\bR}{{\mathbb{R}}}

\newcommand{\rL}{{\mathrm{L}}}
\newcommand{\rR}{{\mathrm{R}}}

\newcommand{\mean}[1]{\overline{#1}}

\newcommand{\vol}{\mathop{\rm vol}}
\newcommand{\diag}{\mathop{\rm diag}}
\newcommand{\sign}{\mathop{\rm sign}}

\newcommand{\beq}{\begin{equation}}
\newcommand{\eeq}{\end{equation}}
\newcommand{\beqa}{\begin{eqnarray}}
\newcommand{\eeqa}{\end{eqnarray}}
\newcommand{\beqas}{\begin{eqnarray*}}
\newcommand{\eeqas}{\end{eqnarray*}}

\newcommand{\A}{{\cal A}}
\newcommand{\B}{{\cal B}}
\newcommand{\C}{{\cal C}}
\newcommand{\D}{{\cal D}}

\begin{document}

\title{Nanocoolers}

\author{Martin Horvat}
\address{Department of Physics, Faculty of Mathematics and Physics, University of Ljubljana, Jadranska 19, SI-1000 Ljubljana, Slovenia}
\eads{\mailto{martin.horvat@fmf.uni-lj.si}}

\author{Toma\v z Prosen}
\address{Department of Physics, Faculty of Mathematics and Physics, University of Ljubljana, Jadranska 19, SI-1000 Ljubljana, Slovenia}
\eads{\mailto{tomaz.prosen@fmf.uni-lj.si}}

\author{Giulio Casati}
\address{CNISM, CNR-INFM \& Center for Nonlinear and Complex Systems, Universita` degli Studi dellÕInsubria, Via Valleggio 11, 22100 Como, Italy, \\
Istituto Nazionale di Fisica Nucleare, Sezione di Milano, Via Celoria 16, 20133 Milano, Italy}

\eads{\mailto{giulio.casati@uninsubria.it}}

\date{\today}

\begin{abstract}

We present a simple kinematic model of a non-equilibrium steady state device, which can operate either as a heat engine or as a refrigerator. The model is composed of two or more scattering channels where the motion is fully described by deterministic
classical dynamics, which connect a pair of stochastic (infinite) heat and particle baths at unequal temperatures. We discuss precise kinematic conditions under which our model may approach Carnot's optimal efficiency in different situations.

\end{abstract}


\submitto{{\it J. Stat. Mech.}}


\section{Introduction}

It has been recognized in recent years that an efficient energy conversion and transformation between heat and work on all scales, in particular at the nanoscale, poses one of the main challenges for future social needs. One of the most promising technologies is a thermoelectric circuit which can simply convert a heat flux from a hot to cold place into an electric power, or conversely, using the electric power to pump heat from cold to hot place \cite{mahan97}. However, although such devices have been known for a long time, and even their technological applications have been diverse for several decades, their thermodynamic efficiency remains quite low compared to compressor based refrigerators or fuel burning internal combustion engines. One of the main reasons for such low efficiency is a  rather poor theoretical understanding for the relatively low figure of merit, the so called ZT in all known materials at room temperature.

Recently, a dynamical system approach has been proposed \cite{casati08, saito10} in an attempt to better understand the mechanism which could lead to an increase of ZT and thus of thermoelectic efficiency. Along these lines we propose here a mathematically simple, general model of a steady state thermoelectric device, which can be analyzed theoretically within the framework of classical statistical mechanics. The model is composed of two or more classical two-port scattering channels where the motion (of non-interacting charged particles) is fully deterministic and energy conserving, and which connect two stochastic, infinite heat and particle baths at different temperatures. The motion in the longitudinal direction connecting the two baths is quasi-free except for point scatterers in the middle of the channels, which have certain simple properties. Namely the transmission only depends on the (kinetic) energy of the particle. It is argued that the dependence on energy of the transmission (scattering) function is crucial in order to obtain non-vanishing efficiency. The particles are only subjected to a bias voltage that is used to input or output the work from the system.

The model we discuss here, even quite abstract, is the most general case of a self-contained (compact) classical heat engine model, composed of two thermo-chemical baths with non-interacting particles. Its simplicity allows theoretical analysis and therefore a better understanding of how a thermo-electric engine operates and the mechanism underlying the behavior of ZT. Indeed in our paper we develop an easy-to-follow formalism in order to discuss particle and heat currents inside the model as well as its power and thermoelectric efficiency. The model's simplicity indicates that engines based on its principles could be practically built and so the model may be important for future design of heat engines.

More precisely we derive analytical expressions for the engine's efficiency, power and the parameters that optimize efficiency (section 2). In the linear temperature difference regime (section 3) in particular, there are cases like window-type scattering functions (section 5) that can be worked out exactly. It is particularly interesting that in the linear temperature difference regime, refrigeration and power generation modes of operation can be treated symmetrically.
Using our formalism, it is possible to discuss the precise conditions under which the model attains Carnot's efficiency. We show in section 4, that a generalized figure of merit is essentially given as a ratio of the variance of Seebeck coefficients for various channels divided by the average variance of the transmitted particle's energy. It is found that  {\em narrow energy filtering} is a good mechanism for increasing thermoelectric efficiency of the engines based on our classical model. Similar effect was previously observed in quantum models, see e.g. \cite{maham96}.
In section 6 we worked out semi-analytically and numerically a slightly different model of thermoelectric engine where the scattering is provided by a homogeneous transverse magnetic field. In this setup, power and efficiency turn out to be tunable via the external magnetic field; however the maximal efficiency remains rather low.

Some more technical aspects of our formalism which are not essential for the main results are presented in two appendices. In appendix A we present consistent construction of classical scattering matrices and their concatenation for structured scattering channels.
While most of the work in this paper refers to the steady-state operation of heat-engine models, we dedicate appendix B to the discussion of relaxation times to the steady state. In particular we show that non-vanishing bias voltages in all channels render strictly finite relaxation times.

As discussed in the following sections, our formalism allows an efficient numerical as well as analytical treatment of concrete not trivial cases of classical-mechanical models of heat engines and helps us to identify the main transport properties controlling power generation and efficiency.

\section{Construction of the heat engine or refrigerator}
  
\begin{figure}[!htb]
\centering
\includegraphics[width=8cm]{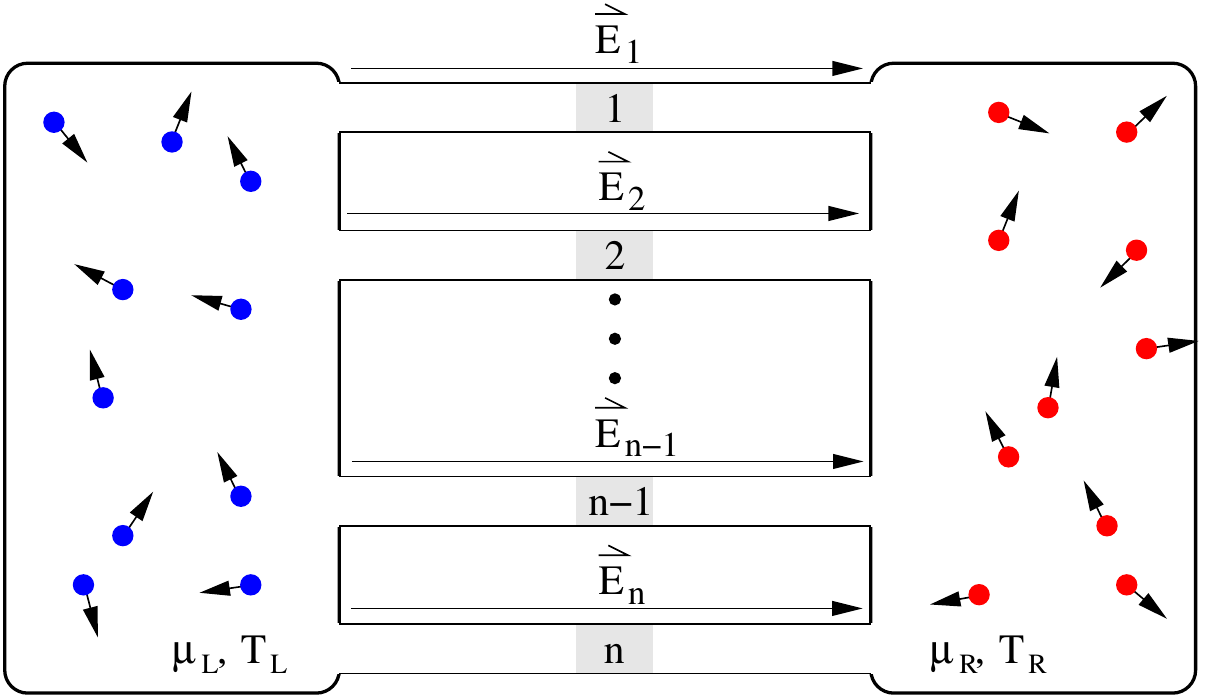}
\caption{Schematic figure of the heat engine.}
\label{pic:multi_channel_scheme}
\end{figure}

The heat engine or refrigerator is constructed of two $d$-dimensional thermo-chemical baths connected by two or more channels as depicted in figure \ref{pic:multi_channel_scheme}. The particles in the system have $d$ degrees of freedom, equal positive charge $e$ and equal mass $m$. We use units in which the charge $e$, mass $m$ and Boltzmann constant $k_{\rm B}$ are equal to unity, $m=e=k_{\rm B}=1$. Along the direction of the channels we introduce bias electrical fields $E_i$, which are described by bias voltages $U_i$ between the left and right baths. These voltages are used to extract work (or power) from the system in the case of power producing heat engine or to supply the work in the case of the refrigerator. The Hamiltonian $H$ of a particle with momentum $p=(p_1,\ldots,p_d)\in\bR^d$ and position $q=(q_1,\ldots,q_d)\in\bR^d$ inside the $i$th channel is given by 
$$
  H(q,p) = W(p) + U_i q_1 \>,\qquad  W(p)= \frac{1}{2} |p|^2\>.
$$
The coordinate axis $q_1$ runs longitudinally along the channel with units chosen so that $q_1 \in [0,1]$, where $q_1=0$ and $q_1=1$ mark the left and the right end of the channel, respectively.
Let there be $n$ channels connecting the left bath side ($\nu=L$) with the right bath side ($\nu=R$) via the holes $\Gamma_{i} \subset \bR^{d-1}$ perpendicular (transverse) to the direction of the channel of ``area" $A_{i} = \vol \Gamma_{i}$. For convenience we assume that the holes through which the channel $i$ is attached to the left and the right bath are represented by identical sets $\Gamma_i$ (analysis could easily be generalized if necessary). The total area of the holes is $A =\sum_{i=1}^n A_{i}$. The ratio $r_{i}=A_{i}/A$ represents the probability that a particle effused from the bath goes into $i$-th channel, from either side. The dynamics in each channel is reversible and energy conserving. The transmission function for a particle from $\nu$-side over $i$th channel is defined as
\beq
 \!\!\!\!\!\!\!\!\!\!\!\!\!\!\!\!\!\!\tau_{\nu,i} (x,p, U_i) = 
  \left\{
  \begin{array}{lll}
   1 &:& {\textrm{\small particle is effused at $x\in\Gamma_i$ with momentum $p\in \bR^d$ from side} \atop \textrm{\small $\nu$ and transmitted through channel $i$ with bias voltage $U_i$}}\\
   0 &:& \textrm{particle is reflected back into the same bath} 
  \end{array}\right.
  \label{eq:trans_def}
\eeq
A scheme for explicit construction of the transmission functions is discussed in appendix A.
In the left (right) bath the particles are at temperature $T_\rL$ ($T_\rR$) and chemical potential $\mu_\rL$ ($\mu_\rR$) and are effused into $i$-th channels with the rate $r_i \gamma_\rL$ ($r_i \gamma_\rR$). The effusion rates $\gamma_{\rL}$ and $\gamma_{\rR}$ are connected with the chemical potential $\mu_\nu$ and reciprocal temperature $\beta_\nu$ via the formula
$$
  \beta \mu_\nu = 
  \log \left(C \beta_\nu^\frac{d+1}{2} A_\nu \gamma_\nu \right)\>,\qquad  \nu \in {\rL,\rR} \>, 
$$
where $C$ is a constant depending only on the mass of the particles. The particles effused from $\nu$-bath into channels have the momentum $p\in\bR^d$ distributed according to the Boltzmann distribution
\beq
 P(p,\sigma,\beta)
  = \beta \left( \frac{\beta}{2\pi} \right)^{\frac{d-1}{2}} |p_1|
  \exp\left(-\frac{1}{2}\beta |p|^2\right) \theta(\sigma p_1)\>,
  \label{eq:distr_effusion}
\eeq
with unit step function $\theta(x) = (0 : x \le 0 ; 1 :{\rm otherwise})$. Let us now  introduce the probability $t_{\nu,i}$ that a particle effused from side $\nu$ is transmitted over $i$-th channel, and the average particle kinetic energy $q_{\nu,i}$, which are computed as
\beq
  (t_{\nu,i}, q_{\nu,i}) =  
  \frac{1}{A_{ i}} \int_{\bR^d} \dd p \, P(p,\sigma_\nu,\beta_\nu)(1 , W(p))\!\int_{\Gamma_i}\!\dd x\, \tau_{\nu,i}(x,p,U_i) \>,
  \label{eq:gen_tranp_coef}
\eeq
with $\sigma_\rL = 1$, $\sigma_\rR = -1$. Notice that at equal temperatures $T_\rL = T_\rR$ and zero fields $U_i = 0$ we have $t_{\rL,i} = t_{\rR,i}$ and $q_{\rL,i} = q_{\rR,i}$. The particle currents $j_{\rho,i}$ within $i$-th channel and the heat currents $j_{q,i}|_\nu$ \cite{degroot} exchanged between the $\nu$-side bath and $i$-th channel are given by
\begin{eqnarray}
  j_{\rho,i} 
  = r_i (\gamma_\rL t_{\rL,i} - \gamma_\rR t_{\rR,i})\>,
  \label{eq:part}\\
  j_{q,i}|_\rL 
  = r_i\left( \gamma_\rL q_{\rL,i} - \gamma_\rR (q_{\rR,i} + t_{\rR,i} U_i)\right)\>, 
  \label{eq:heatL}\\
  j_{q,i}|_\rR 
  = r_i\left( \gamma_\rL (q_{\rL,i} - t_{\rL,i} U_i) - \gamma_\rR q_{\rR,i}\right)\>.
  \label{eq:heatR}
\end{eqnarray}
In the stationary state, the engine model should obey two basic principles (compatible, or even equivalent to the 2nd law):
\begin{itemize}
\item The net particle current from one bath to another is zero $\sum_{i=1}^n j_{\rho,i} = 0$.
\item If temperatures of the baths are equal, then the particle current in each channel is zero.
\end{itemize}
Assuming that the average injection rate $\mean{\gamma}= \frac{1}{2} (\gamma_\rL + \gamma_\rR)$ is known (say it is an input parameter of the model) the two above principles yield the relation 
$$
  \frac{\gamma_\rR}{\mean{\gamma}}-1 
  = 1-\frac{\gamma_\rL}{\mean{\gamma}}
  = \frac{\A_\rL - \A_\rR}{\A_\rL + \A_\rR}\>.
$$
which uniquely determine the injection rates $\gamma_\rL$ and $\gamma_\rR$. After introducing auxiliary variables
$$
  (\A_\nu, \B_\nu,\C_\nu) = 
  \sum_{i=1}^n r_i (t_{\nu,i}, t_{\nu,i} U_i, q_{\nu,i})\>,
$$
the stationary particle currents in the channels read
$$
  j_{\rho,i} = 
    \frac{\mean{\gamma} r_i}{\A_\rL +\A_\rR} 
    \left[t_{\rL,i} \A_\rR - t_{\rR,i} \A_\rL \right]\>
$$
and the power produced by the particle flow is given by
$$
  P = \sum_{i=1}^n U_i j_{\rho,i} 
    = \frac{\mean{\gamma} r_i}{\A_\rL + \A_\rR}  
    \left[ \B_\rL \A_\rR - \B_\rR \A_\rL \right] \>.
$$
A non-vanishing power is produced, if and only if $\D = \B_\rL \A_\rR - \B_\rR \A_\rL$ is non-zero. Similarly, we define the total stationary heat current $Q_\nu  = \sum_{i=1}^n j_{q,i}|_\nu$ on both sides of the channels 
\begin{eqnarray}
  Q_\rL 
  =\frac{2\mean{\gamma}}{\A_\rL + \A_\rR} 
   \left[ \C_\rL \A_\rR  - \C_\rR \A_\rL + \B_\rR \A_\rL \right]\>, \\
  Q_\rR 
  =\frac{2\mean{\gamma}}{\A_\rL+\A_\rR} 
   \left[ \C_\rL \A_\rR  - \C_\rR \A_\rL + \B_\rL \A_\rR \right]\>.
\end{eqnarray}
Notice the similar expression for the total heat current on the left and right side, which become identical in the linear temperature-difference regime. We can think of the apparatus as a power producing heat engine or as a refrigerator. In the heat engine regime, the heat is transferred from the hotter bath $\nu$ to the colder bath in such a way that the particles in the channels travel against the bias voltage and generate a power $P > 0$. In the case of the refrigerator regime, the heat is pumped from the colder bath of the side $\nu$ by appropriately setting the potentials in the channels and thereby a power $P < 0$ is used. By introducing a binary index
$$
\zeta = \left \{ 
  \begin{array}{lll} 
  +1 &:& \textrm{in the case of the power producing heat engine mode} \\
  -1 &:& \textrm{in the case of the refrigerator mode}
  \end{array}
\right. \>,
$$
we can write the efficiency $\eta$ for both cases as
\beq
 \eta 
 = \left(\frac{\zeta P}{\sigma_\nu Q_\nu}\right)^\zeta
 \le \eta_{\rm Carnot} 
 = \left(\frac{|T_\rL - T_\rR|}{T_\nu}\right)^\zeta \>,
 \label{eq:eta_gen}
\eeq
with  $\sigma_\nu Q_\nu>0$. The efficiency $\eta$ of the heat engine and refrigerator is bounded from above by the efficiency of the ideal Carnot cycle $\eta_{\rm Carnot}$. For a given setup of scatterers the efficiency can be optimized by choosing appropriate potentials. In general the optimization can be done only numerically, because the expressions for $t_{\nu,i}$ and $q_{\nu,i}$ are usually nonlinear functions of the potentials. Only in the linear temperature-difference regime, discussed in the  next section, the optimization can be done analytically.

\section{Linear temperature-difference regime}

In the previous section we have discussed the general framework of our heat engine model. In the following we elaborate on the linear temperature-difference regime, where the temperature difference $\delta T = T_\rR - T_\rL$ is much smaller than the average temperature $\mean{T}=\frac{1}{2}(T_\rL + T_\rR)$. Consequently, the difference in injection rates $\delta \gamma = \gamma_\rR - \gamma_\rL$ and in electrical potentials $U_i$ on both sides are also small. These conditions permit a perturbative treatment of energy and particle currents and enable an analytic optimization of the efficiency.\par
The idea is to expand the particle (\ref{eq:part}) and the heat currents (\ref{eq:heatL}) (\ref{eq:heatR}), which depend on transport coefficients $t_{s,i}$ and $q_{s,i}$ (\ref{eq:gen_tranp_coef}) and injection rates $\gamma_s$, in terms of $\delta T$, $\delta \gamma$ and $U_i$. First we write the injection rates from both sides
$$
  \gamma_\nu = \mean{\gamma} + \frac{1}{2} \sigma_\nu\delta \gamma,\qquad 
  \nu \in \{\rL,\rR\} \>.
$$
using the difference in injection rates $\delta \gamma$ reading
\beq
  \delta \gamma = \mean{\gamma} \left[ 
    \left(\mean{\mu} - \frac{d+1}{2\beta}\right)\delta \beta + 
    \mean{\beta} \delta \mu
  \right] \>.
  \label{eq:lingamma}
\eeq
where we introduce the average and the difference of chemical potential $\mean{\mu} = (\mu_\rL + \mu_\rR)/2$ and $\delta\mu = \mu_\rR - \mu_\rL$, respectively, and the reciprocal average temperature $\mean{\beta} = \mean{T}^{-1}$. Next we expand the transport coefficients $t_{s,i}$ and $q_{s,i}$ linearly in $\beta$ and $U_i$ around the equilibrium point $\beta=\mean{\beta}$, $U_i=0$:
\begin{eqnarray}
  t_{\nu,i} &=& 
    t_{\nu,i}|_0 + 
    \partial_\beta t_{\nu,i}|_0\, \delta \beta +
    \sigma_\nu\partial_{U_i} t_{\nu,i}|_0\, U_i\\
  q_{\nu,i} &=& 
    q_{\nu,i}|_0 + 
    \partial_\beta q_{\nu,i}|_0\, \delta \beta +
    \sigma_\nu\partial_{U_i} q_{\nu,i}|_0\, U_i\>.
\end{eqnarray}
Then, by using the transport coefficients at the equilibrium point
\beq
\!\!\!\!\!\!\!\!\!\left(t_i^{(0)}, q_i^{(0)}, k_i^{(0)}\right) =  
   \frac{1}{A_i} \int_{\bR^d} \dd p\, P(p,1,\mean{\beta})
  (1 , W(p) ,W(p)^2)\!\int_{\Gamma_i}\!\dd x\, \tau_{\nu,i}(x,p,0) \>,
  \label{eq:coef_gen}
\eeq
we may write the temperature expansion coefficients as
$$
  \partial_\beta t_{\nu,i}|_0 
  = \frac{d+1}{2\mean{\beta}} t_i^{(0)} - q_i^{(0)}\>,\qquad 
  \partial_\beta q_{\nu,i}|_0 
  = \frac{d+1}{2\mean{\beta}} q_i^{(0)} - k_i^{(0)} \>.
$$

The latter depend only on the mean reciprocal temperature $\mean{\beta}$. For the time being we assume that the channels are straight cylindrical leads with a local scatterer in the middle (i.e. located in the center of each channel) and that the scattering process depends only on the kinetic energy, i.e.
\beq
  \tau_{\nu,i}(x,p,U_i):= 
  \psi_i\left(W(p) -\frac{1}{2} \sigma_\nu U_i\right) 
  \theta\left(\frac{1}{2} p_1^2-\max(0,\sigma_v U)\right) \>,
  \label{eq:onshell_def}
\eeq
where $\psi$ is a transmission function at fixed energy of the scatterer inside the channel. We refer to $\psi$ as the on-shell transmission function. It is possible to show that 
\beq
  \partial_{U_i} t_{\nu,i}|_0 
  = -\frac{1}{2} \sigma_\nu \mean{\beta} t_i^{(0)}\>,\qquad
  \partial_{U_i} q_{\nu,i}|_0 
  = -\frac{1}{2} \sigma_\nu (\mean{\beta} q_i^{(0)}- t_i^{(0)})\>.
 \label{eq:lin_cond}
\eeq
and the transport coefficients at equilibrium point simplify to
\beq
  \left(t_i^{(0)}, q_i^{(0)}, k_i^{(0)}\right) =
  \frac{\mean{\beta}^{\frac{d+1}{2}}}{ \left(\frac{d-1}{2}\right)!}
  \int_0^\infty \dd W\,
  W^{\frac{d-1}{2}}
  e^{-\mean{\beta}W} \psi_i(W) (1, W, W^2) \>.
  \label{eq:coef_energy}
\eeq
The relations (\ref{eq:lin_cond}) are valid for scatterers of more general type even though it might be difficult to give a rigorous proof. In order to simplify the following discussion we introduce auxiliary constants
\beq
  t_i = \mean{\gamma} \mean{\beta} r_i t_i^{(0)}\>,\qquad
  q_i = \mean{\gamma} \mean{\beta} r_i q_i^{(0)}\>,\qquad
  k_i = \mean{\gamma} \mean{\beta} r_i k_i^{(0)}\>.
  \label{eq:coef_lin}
\eeq
and the relative reciprocal temperature difference 
$$
  \xi = \delta \beta/\mean{\beta}.
$$  
Taking into account the linear expansion of transport coefficients (\ref{eq:lin_cond}) and injection rates (\ref{eq:lingamma}) we can write the particle currents $j_{\rho,i}$  (\ref{eq:part}) and  the heat currents $j_{q,i}$ (\ref{eq:heatL}) as
\beqa
   j_{\rho,i} &=& - t_i(U_i + \delta\mu) + \xi q_i\>,\\
   j_{q,i} &=& -q_i(U_i + \delta\mu) + \xi k_i\>.
\eeqa
where we use the fact that the heat current on the left and right side are equal, $j_{q,i} := j_{q,i}|_\rL = j_{q,i}|_\rR$, in the considered regime. Without loss of generality, the mean value of the chemical potential $\mean{\mu}$ is set to zero, as only differences are important. The expression for the currents can be given in the matrix form
\beq
  \left[\begin{array}{c} 
    j_{\rho,i} \\ 
    j_{q,i} 
  \end{array}\right] =
  \left [
    \begin{array}{cc}
     t_i & q_i \\
     q_i & k_i
    \end{array}
  \right ]
  \left[\begin{array}{c}
  \xi\\ 
  -(U_i +\delta \mu)
  \end{array}\right]\>,
  \label{eq:currmat} 
\eeq
showing that they indeed satisfy the Onsager reciprocity relations \cite{degroot}. It is convenient to express $q_i$ and $k_i$ in terms of positive quantities $S_i = q_i/t_i$ and $K_i = k_i - t_i S_i^2$ and introduce $n$-component column vectors 
$$
 t = [t_i]_{i=1}^n\>,\quad
 S = [S_i]_{i=1}^n\>,\quad
 K = [K_i]_{i=1}^n\>,\quad
 U = [U_i]_{i=1}^n\>,\quad
 \one = [1,\ldots,1]^T .
$$
The stationary condition of zero net current  $\sum_{i=1}^n j_{\rho,i}=0$ determines the difference $\delta\mu$ of chemical potentials
$$
 \delta \mu =  \frac{\xi t^T S - t^T U}{\one^T t}\>.
$$
By using this result we write the stationary power $P= \sum_i j_{\rho, i} \delta U_i$ and the stationary net heat current $Q  = \sum_i j_{q, i}$ as
$$
  P = (\xi S - U)^T M U\>,\qquad
  Q = \xi \left[ S^T M S  + (\one^T K)  (\one^T t) \right] - M S\>.
$$
where the matrix $M= (\one^T t) \diag(t) - t t^T$. Notice that $P$ and $Q$ are invariant under the transformation $U \to U + {\rm const}$. The efficiency $\eta$  (\ref{eq:eta_gen}) of the heat engine or refrigerator can be optimized w.r.t. to the field vector $U$ by solving the set of equations $\partial \eta / \partial U = 0$. Let us introduce a quantity 
\beq
  y = \frac{S^T M S}{(\one^T K) (\one^T t)} > 0
  \label{eq:y}
\eeq
called the merit of efficiency. Then the optimal efficiency $\eta^*$ of the heat engine or the refrigerator is equal to
\beq
  \eta^* = \eta_{\rm Carnot} \frac{\sqrt{1+y}-1}{\sqrt{1+y}+1}\>,
  \label{eq:eta_opt}
\eeq
and the corresponding vector of the bias voltages $U^*$ and the output power $P^*$  are written as
\beqa
  U^* = \xi  \frac{\sqrt{1+y}}{y} \left(\sqrt{1+y} - \zeta \right)
  \left ( S - \frac{1}{n} (\one^T  S)\one \right) \>,\\
 P^* = 
 -\left(\frac{\xi}{y}\right)^2 (S^T M S) \sqrt{1+y} 
 \left(2 \sqrt{1+y} - \zeta(2 + y)\right)\>,\\
 \label{eq:power_opt}
 Q^* = 
 \zeta \frac{\xi}{y} (S^T M S) \sqrt{1+y}\>,
\eeqa
where we use for Carnot efficiency $\eta_{\rm Carnot} = \zeta \xi^{\zeta}$ (\ref{eq:eta_gen}). The relative efficiency $\eta/\eta_{\rm Carnot}$ of the heat engine or refrigerator is a monotonically increasing function of $y$. In the limit $y\to\infty$ we reach the Carnot cycle efficiency. In the limit of two channels $n=2$ we obtain the expression already derived in \cite{jiao, horvat08}. In this case the merit of efficiency reads
$$
  y = \frac{(S_1- S_2)^2}{(K_1 + K_2) (1/t_1 + 1/t_2)}\>.
$$

\section{Optimally efficient engines}

The efficiency of a steady state device that is working as a heat engine or refrigerator can be far from maximal, Carnot cycle efficiency. Here we describe a general strategy to increase efficiency which is based on our model in the linear temperature difference regime.\par
Let us consider a $n$-channel heat engine or refrigerator with scattering depending only on the kinetic energy $\tau_{\nu,i}(x,p,0) = \psi_i(W(p))$, at reciprocal average temperature $\mean{\beta}$. We want to determine which properties of the scatterers may lead to maximal efficiency. By taking into account the transmission function and the distribution of momenta by the  effusion process (\ref{eq:distr_effusion}) we know that the particle's energy $W$ transmitted over $i$th channel has the probability density
$$
  P_i(W) = 
  \frac{\mean{\beta}^{\frac{d+1}{2}}}
       {t_i^{(0)}\left(\frac{d+1}{2}\right)!}
  e^{-\mean{\beta}W} W^{\frac{d-1}{2}} \psi_i(W)\>.
$$
with average $\mean{W}_i= S_i$ and variance $V_i := \mean{W_i^2}-\mean{W_i}^2=K_i/t_i$.  It is interesting to notice that in our notation the figure of merit ZT \cite{casati08} of the $i$th channel is given simply by $S_i^2/V_i$.\par
In order to further simplify the discussion we assume that the bath openings are equal, $r_i =: r$ and that the scatterers have equal transmission probability $t_i =: t$. This makes the channels comparable in the ability to transmit particles.  In this case the merit of efficiency of our engine is simplified to
\beq
 y = \frac{\sum_i S_i^2 - \frac{1}{n}(\sum_i S_i)^2}{\sum_i V_i} \>.
 \label{eq:eta_special}
\eeq
Imagine now that the scatterers act as energy window filters. Then by narrowing the window, the variances $V_i$ will decrease and the averages $S_i$ will converge to some definite value. If the numerator in (\ref{eq:eta_special}) does not vanish, then the merit of efficiency $y$ will diverge and consequently 
\beq
  \lim_{V_i\to 0} \eta^* = \eta_{\rm Carnot} \>.
  \label{eq:carnot_limit}
\eeq
We conclude that efficiency of heat engines or refrigerators based on our model can be improved by strongly narrowing the energy window of transmission at the price of smaller  power output $P^*$ or heat current $Q^*$, depending of working regime. The idea of using narrow energy windows to improve thermoelectric efficiency is not new \cite{maham96}. However it is interesting that such principle has a quite general validity.

\section{Engine with a window-shape scatterers}\label{sec:1dwindow}

\begin{figure}[!htb]
\centering
\includegraphics[width=10cm]{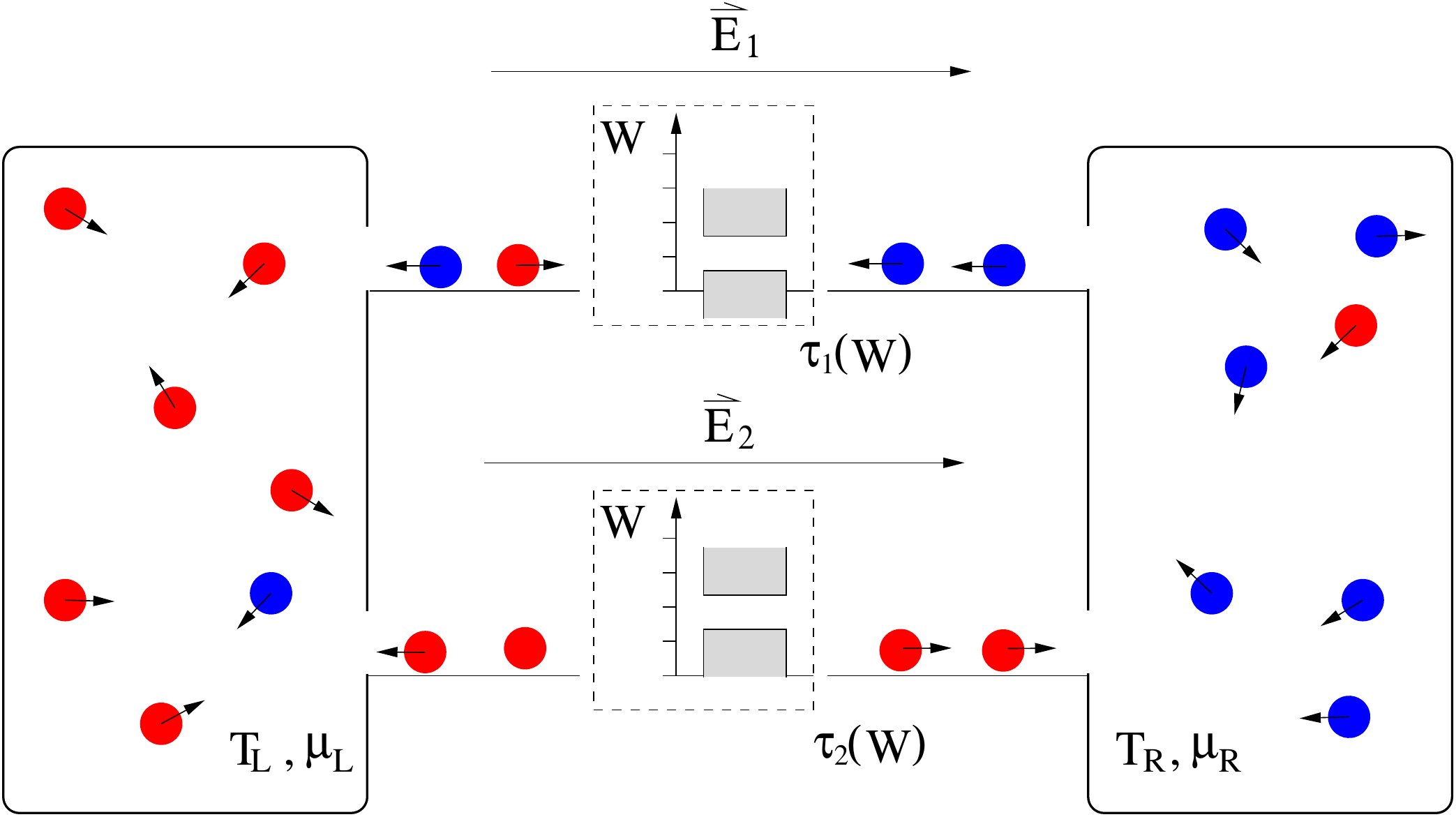}
\caption{Schematic picture of the engine with the channel scatterers in behaving as energy windows.}
\label{pic:2channel}
\end{figure}

Here we study a specific model of an engine (heat engine and refrigerator) with two ($n=2$) one-dimensional ($d=1$) channels with equal relative effusion probabilities $r_1=r_2=\frac{1}{2}$. The particles are effused into a channels from side $\nu$ with momentum $p\in\bR$ distributed with probability density 
$$
   P_\nu (p) = \beta |p| e^{-\frac{1}{2}\beta p^2} \theta(\sigma_\nu p) \>.
$$
In the middle of each channel there is a point-like scatterer, which allows transmission only of particles with kinetic energy in a specific interval and reflects all others. The on-shell transmission function of the scatterer $\psi_i$ (\ref{eq:onshell_def}) in $i$th channel has a window shape in the energy axis that is given by
$$
  \psi_i(W)= \theta(\chi_i + \Delta \chi_i - W) \theta(W - \chi_i) \>,
$$
where $\chi_i$ is the beginning of the energy window and $\Delta \chi_i$ is its width. This case is schematically depicted in figure \ref{pic:2channel}.\par
The dynamics of particles in channels is conservative and symmetric w.r.t. time reversal. Due to simplicity of dynamics the transmission function of the channel (\ref{eq:trans_def}) can be written explicitly and reads
$$
  \tau_{\nu,i} (x,p, U_i) = \psi_i(\frac{1}{2} p^2 - \frac{1}{2} \sigma_\nu U_i)\>.
$$
In the presence of electric field, particles escape from the channels in a finite time of order of magnitude $O(({\min_i|U_i|})^{-1/2})$. On the same time scale the engine relaxes into a non-equilibrium stationary state. \par
\begin{figure}[!htb]
\centering
\rotatebox{90}{\hspace{17mm}\footnotesize $\Delta \chi=\frac{1}{2}$}\hskip5mm%
\includegraphics*[width=5.3cm,bb=0 35 239 227]{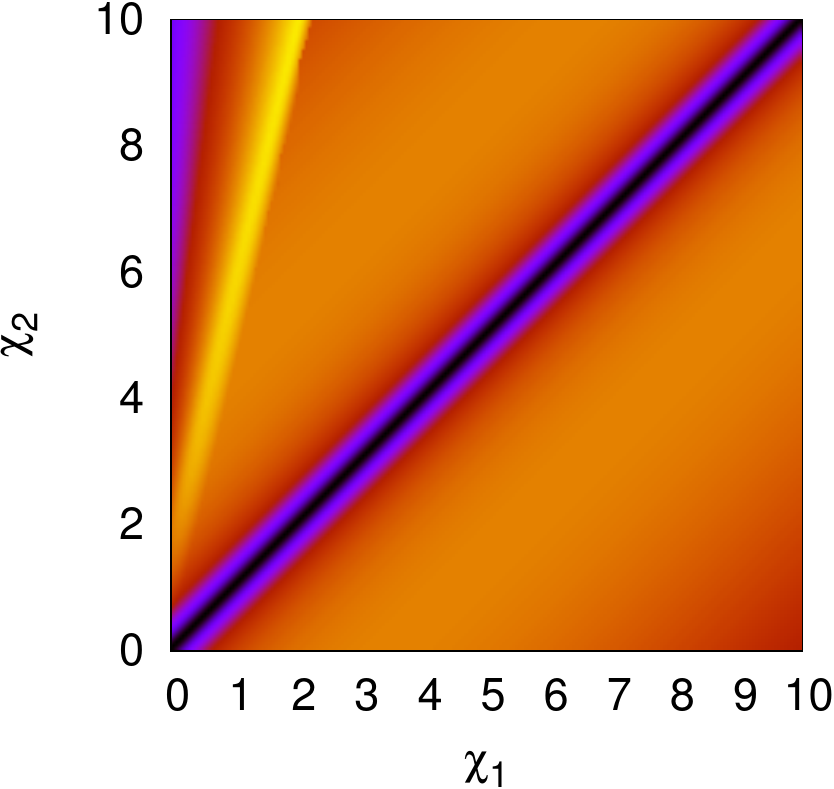}\hskip2pt%
\includegraphics*[width=4.3cm,bb=45 35 239 227]{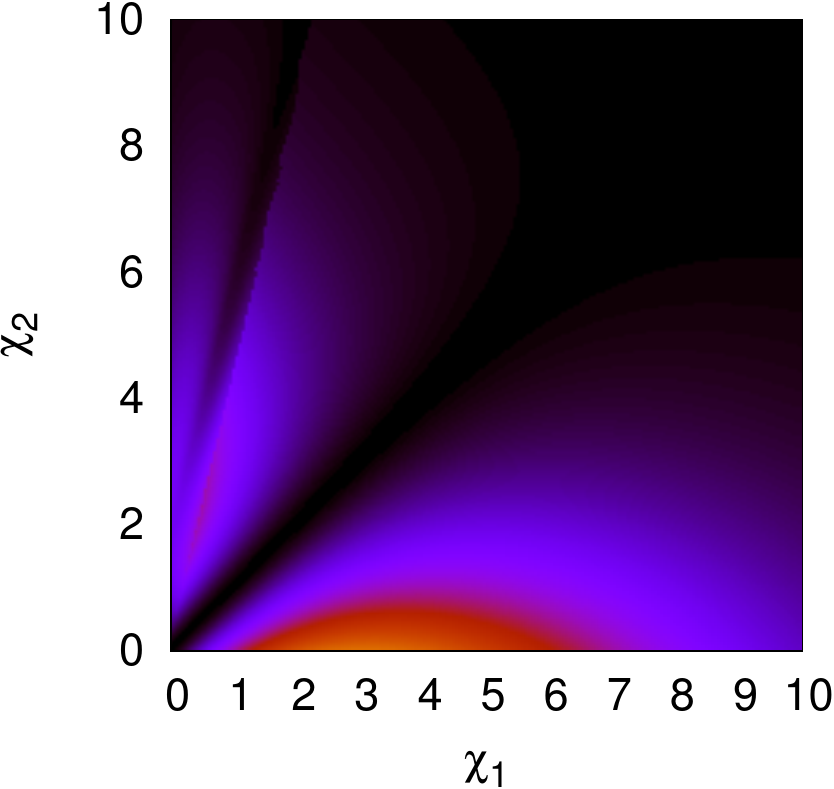}\\
\rotatebox{90}{\hspace{2cm}\footnotesize $\Delta \chi=\frac{1}{4}$}\hskip5mm%
\includegraphics*[width=5.3cm,bb=0 0 239 227]{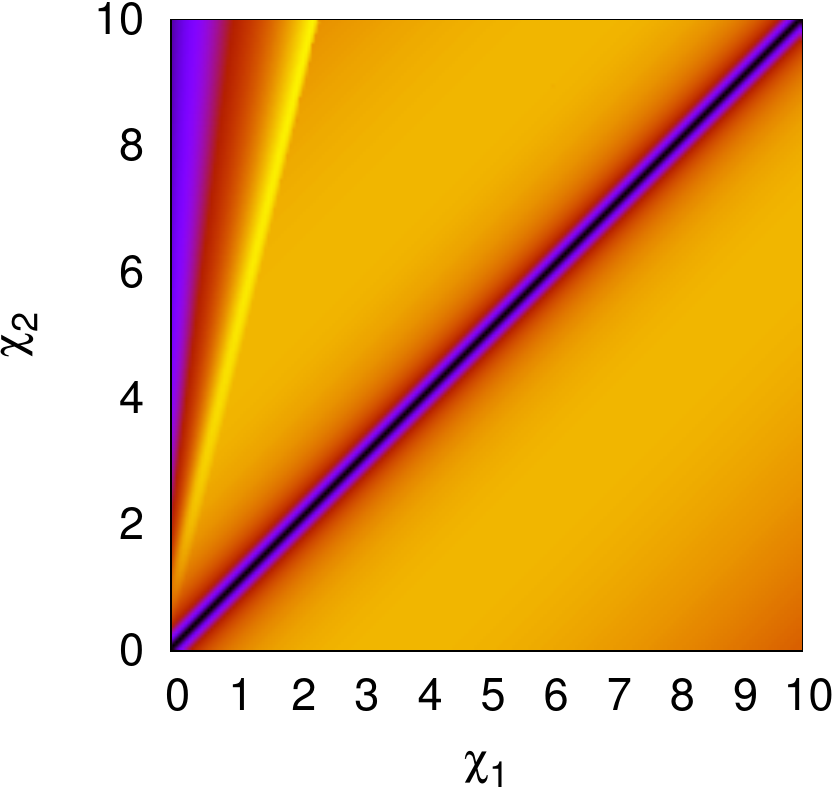}\hskip2pt%
\includegraphics*[width=4.3cm,bb=45 0 239 227]{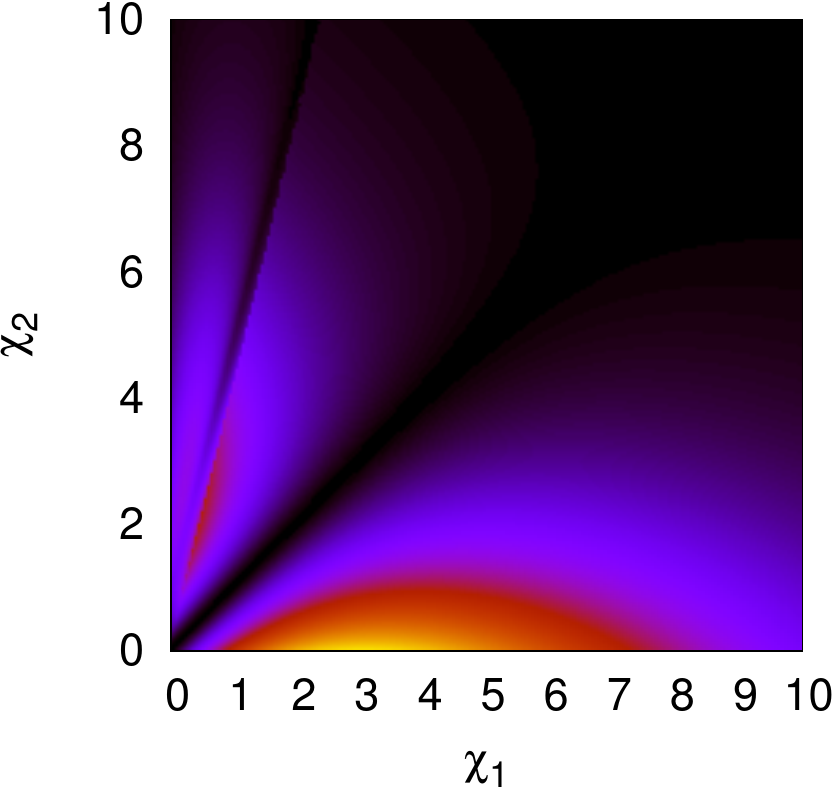}\\[2mm]
\hskip1.8cm%
\includegraphics[width=8.8cm]{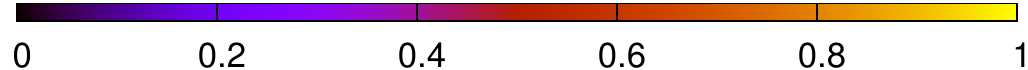}\\
\hskip3cm\hbox to10cm{\footnotesize\hfil (a)\hfil\hfil (b) \hfil}
\caption{The relative optimal efficiency $\eta^*/\eta_{\rm Carnot}$ (a) and the corresponding  rescaled power $a\, P^*/[\mean{T}(\eta_{\rm Carnot}\Delta \chi)^2]$ (b), where for convenience we take $a=13$,  of the heat engine at temperatures $T_\rL=1$ and $T_\rR=2$  ($\mean{T}=1$) and $\mean{\gamma}=1$ for two different width of energy windows $\delta \chi$. Here $\eta_{\rm Carnot}=\frac{1}{2}$.}
\label{pic:win_heat}
\end{figure}
\begin{figure}[!htb]
\centering
\rotatebox{90}{\hspace{17mm}\footnotesize $\Delta \chi=\frac{1}{2}$}\hskip5mm%
\includegraphics*[width=5.3cm,bb=0 35 239 227]{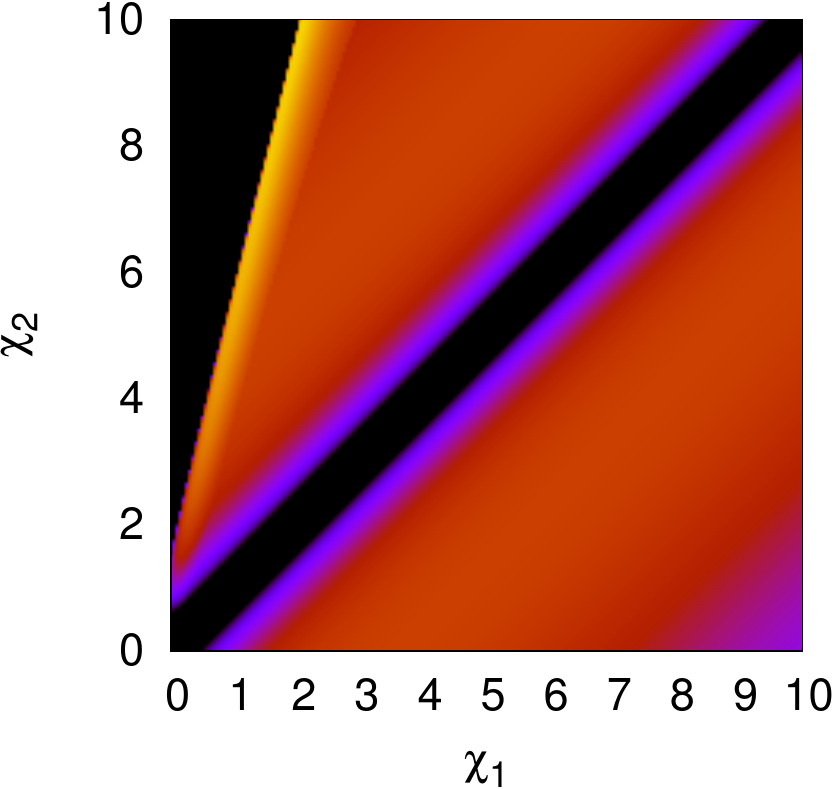}\hskip2pt%
\includegraphics*[width=4.3cm,bb=45 35 239 227]{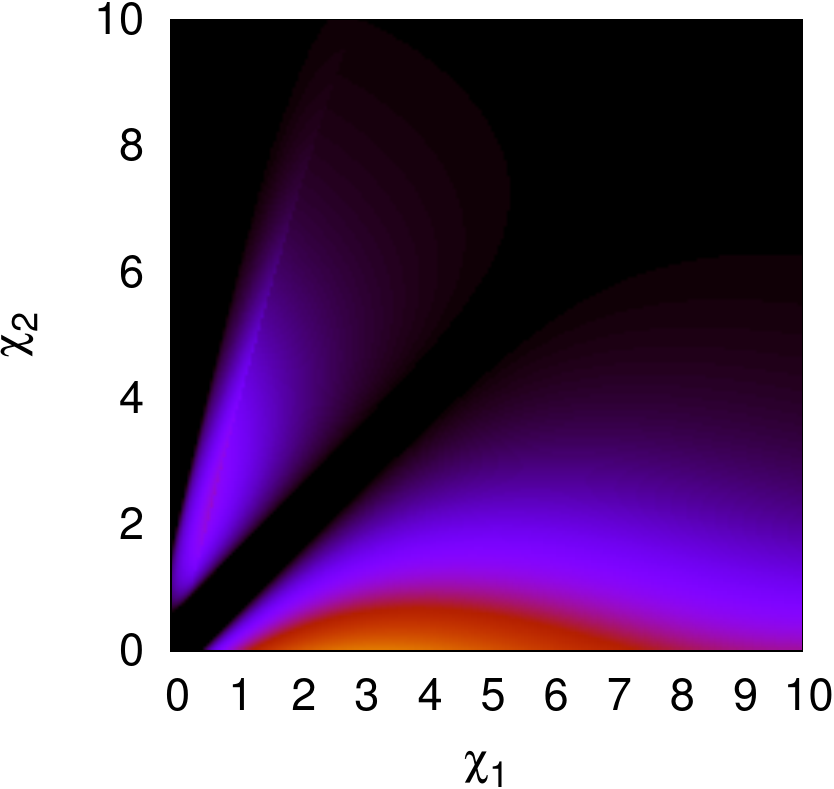}\\
\rotatebox{90}{\hspace{2cm}\footnotesize $\Delta \chi=\frac{1}{4}$}\hskip5mm%
\includegraphics*[width=5.3cm,bb=0 0 239 227]{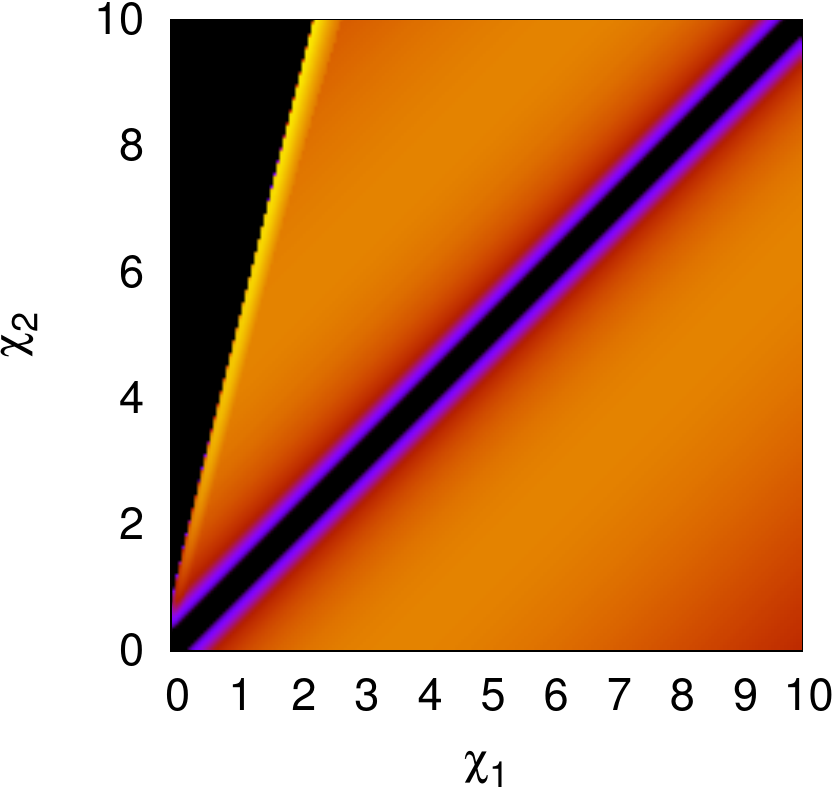}\hskip2pt%
\includegraphics*[width=4.3cm,bb=45 0 239 227]{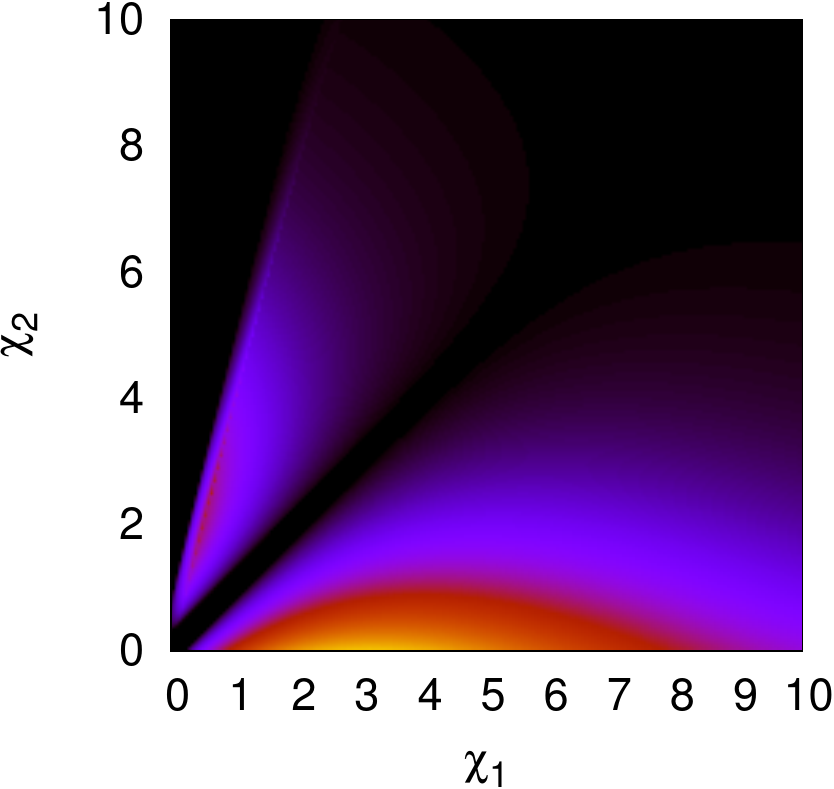}\\[2mm]
\hskip1.8cm%
\includegraphics[width=8.8cm]{figs/heat_win_eta_colorbox.pdf}\\
\hskip3cm\hbox to10cm{\footnotesize\hfil (a)\hfil\hfil (b) \hfil}

\caption{Refrigerator relative optimal efficiency $\eta^*/\eta_{\rm Carnot}$ (a) and corresponding heat current $45\,Q^*/(\mean{T}\eta_{\rm Carnot}^2\Delta \chi)$ (b). Here $\eta_{\rm Carnot}=1$.  For details see the caption under figure \ref{pic:win_heat}.}
\label{pic:win_refrig}
\end{figure}

The performance of such engines can be evaluated analytically but optimization, in general, can only be done numerically. In the following we study engines in which scatterers have transmission functions of the same width $\Delta \chi$. In particular, we set the electric field in the second channel to zero and optimize efficiency by fine-tuning the electric field in the first channel. In figures \ref{pic:win_heat} and \ref{pic:win_refrig} we show the optimal performances of the heat engine and the refrigerator, respectively, in the nonlinear temperature regime as a function of $\chi_i$ for two different $\Delta \chi$.
In figure \ref{pic:win_heat} we plot the optimal efficiency $\eta^*(\chi_1, \chi_2)$ and the corresponding power $P^*(\chi_1, \chi_2)$, which both turn out to be roughly symmetric functions. The symmetry is broken as we apply the electric field just in the first channel. The optimal efficiency, near to Carnot's, is reached on an area lying parallel to the symmetry line $\chi_1=\chi_2$ and, due to asymmetry in the setup, also along a steeper line $\chi_2/\chi_1 = {\rm const}$. A high value of $\eta^*$ and with simultaneous high power output $P^*$ can be achieved in an area near to the $\chi_1$ axis. Note that the maximal possible efficiency slightly increases by decreasing the width $\Delta \chi$ of the energy window in the scatterers.\par
The refrigerator is constructed similarly as the heat engine. The only difference is that we now apply an electric field to revert the direction of the heat current so that the heat is sucked by the engine at the cold bath $Q|_\rL>0$ (we take $T_L < T_R$), and thereby the bias voltage performs some work ($P<0$). We find the optimal performance of the refrigerator by numerically maximizing the efficiency $\eta=Q|_{\rL}/(-P)$ with the constrains $Q|_{\rL}>0$ and $P<0$. This is not a trivial procedure and can easily fail, if the optimal electric fields are either small or large. In figure \ref{pic:win_refrig} we present the relative optimal efficiency $\eta^*/\eta_{\rm Carnot}$ and the corresponding heat current $Q^*$ as a function of the energy window thresholds $\chi_i$ in the nonlinear temperature regime. The results are shown for two widths $\Delta\chi$ of energy windows. 
We see that the dependence of $\eta^*/\eta_{\rm Carnot}$ and $P^*$ on parameters $\chi_i$ and $\Delta \chi$ is very similar to the case of optimally performing heat engine, as expected from the linear theory. The area, where the optimization fails or where efficiency is negative, is painted black and is located along the $\chi_2$ axis. This show that the refrigerator can not be realized for arbitrary setup of scatterers.\par
In the linear temperature regime we are particularly interested in the limit of narrow energy windows. In this limit we can analytically compute the optimal efficiency $\eta^*$ and the corresponding power $P^*$ and heat current $Q^*$. Let us consider scatterers with energy windows of the same width $\Delta \chi$ and at mean energies $\mean{\chi}_1$ and $\mean{\chi}_2$ in the first and the second channel, respectively. In the discussed limit $\Delta x \to 0$ we may write
\beqa
  P^* &=& \frac{\gamma\xi^2 }{6 \beta} [r_1 e^{-x_1} + r_2 e^{-x_2}]a \Delta x + 
  O[(\Delta x)^3]\>, \label{eq:P_win}\\
  Q^* &=&\frac{\zeta\gamma \xi}{6 \beta} [r_1 e^{-x_1} + r_2 e^{-x_2}]a \Delta x + 
  O[(\Delta x)^4]\>, \label{eq:Q_win}\\
  \frac{\eta^*}{\eta_{\rm Carnot}} &=& 1 - \frac{\Delta x }{a} + O[(\Delta x)^2] \>,
  \label{eq:eta_win}
\eeqa
where we define rescaled window thresholds $x_i := \beta \chi_i$, the rescaled widths $\Delta x := \beta \Delta \chi$, and a constant 
$$
 a= \frac{\sqrt{3} |x_2 - x_1|}
    {2 \cosh \left[\frac{1}{2} \left(x_2 - x_1 + \log(\frac{r_1}{r_2})\right)\right]}
%
$$
It is clear that by decreasing the rescaled width $\Delta x$ to zero, the efficiency $\eta$ (\ref{eq:eta_win}) increases linearly up to Carnot's and the leading order of $P^*$ (\ref{eq:P_win}) and $Q^*$ (\ref{eq:Q_win}) decrease quadratically to zero. 

Such behavior of efficiency is expected and agrees with the result (\ref{eq:carnot_limit}) and with the energy filtering mechanism discussed in the context of quantum mechanics \cite{maham96, linke2}.

\section{Heat engine with magnetic fields}

\begin{figure}[!htb]
\centering
\includegraphics[width=10cm]{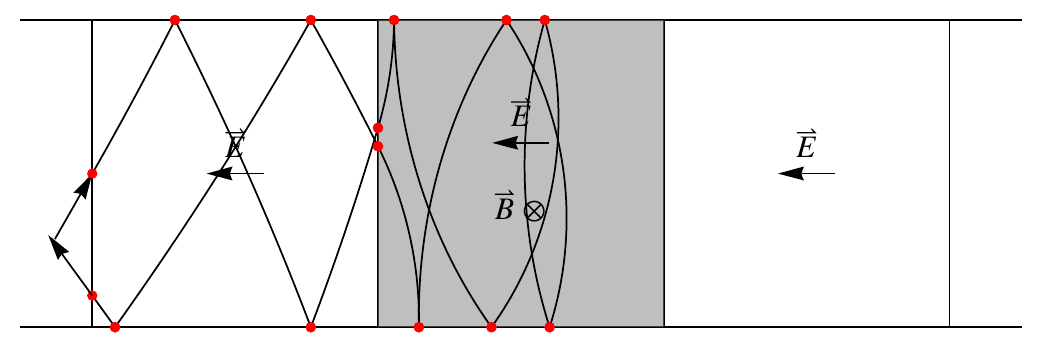}\hskip2pt%
\includegraphics[width=10cm]{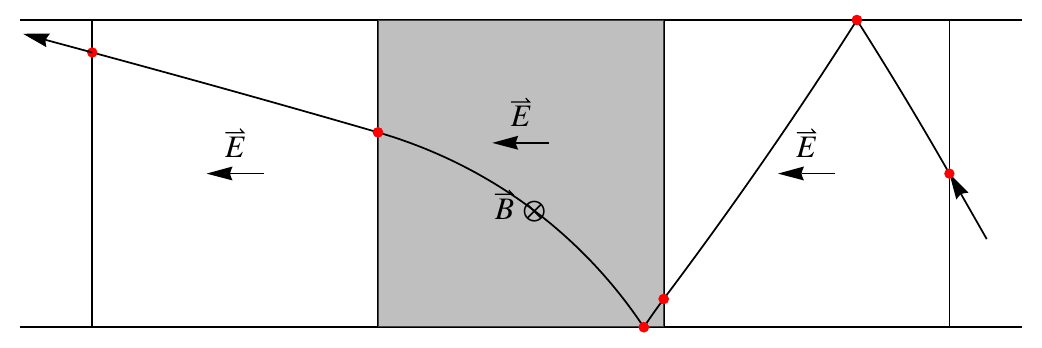}
\caption{An illustration of trajectories in a channel with the middle section immersed into a transverse magnetic field. The lower panel shows the case of a trajectory effused from the other side.}
\label{pic:mag_chan}
\end{figure}

In this section, we consider a slightly different setup where time reversal invariance of the scattering mechanism is broken. We again take a model with two channels that are two dimensional $d=2$ straight leads with hard walls. The $i$th channel is divided into three sections of length $a_i$, $b_i$ and $c_i$ (listed from left to right), see figure \ref{pic:mag_chan}.  The channel's length is $L=a_i+b_i+c_i$ and its width is $h_i$. The magnetic field $B_i$ is applied to the middle section in the $z$ direction and is described by the ``synchrotron orbital frequency"
$$
  \Omega_i = \frac{e B_i}{m} \>,
$$
where $e$ and $m$ are the charge and the mass of the particle, respectively. In addition to the magnetic field, we have a homogeneous electric field $E_i$ along the channel with voltage $U_i=-E_i L_i$ between the ends.
The dynamics in the channels is energy conserving, but not symmetric w.r.t. time reversal.  We may decompose the motion into bounces between channel boundaries of different sections that we may think of as nodes in a scattering graph. Then the scattering of particles is just a way to move in the scattering graph. In the considered case the scattering graph is infinite, due to possibility of very small loops on the boundaries. However, the total length and time spent in these loops is proportional to the length of the boundaries and therefore is finite. Consequently, the particles in the presence of the electric field are ejected from the channels in a finite time. \par
The dynamics of particles in the lead has certain symmetry. In the absence of electric field each trajectory starting on the left side has a counterpart starting on the right side obtained by reflecting the trajectory through the center of the middle section. This means that at $E_i=0$ and with equal bath temperatures $T_{\rm L}=T_{\rm R}$ we have $t_{\rL, i}=t_{\rR,i}$ and $q_{\rL, i}=q_{\rR, i}$.
The transport coefficients $t_{\nu, i}$ and $q_{\nu, i}$ depend only on the absolute value of the magnetic field. Indeed change in the sign of the magnetic field just mirrors the dynamics around the middle of the channel.\par
In the following we optimize the efficiency as a function of the electric field at the fixed orbital frequencies $\Omega_i>0$. In numerical experiments the channels are taken of equal width, implying $r_1=r_2=\frac{1}{2}$, and with the unit average injection rate $\mean{\gamma}=1$.

\subsection{Linear temperature-difference regime} 

\begin{figure}[!htb]
\centering
\includegraphics[width=7.5cm]{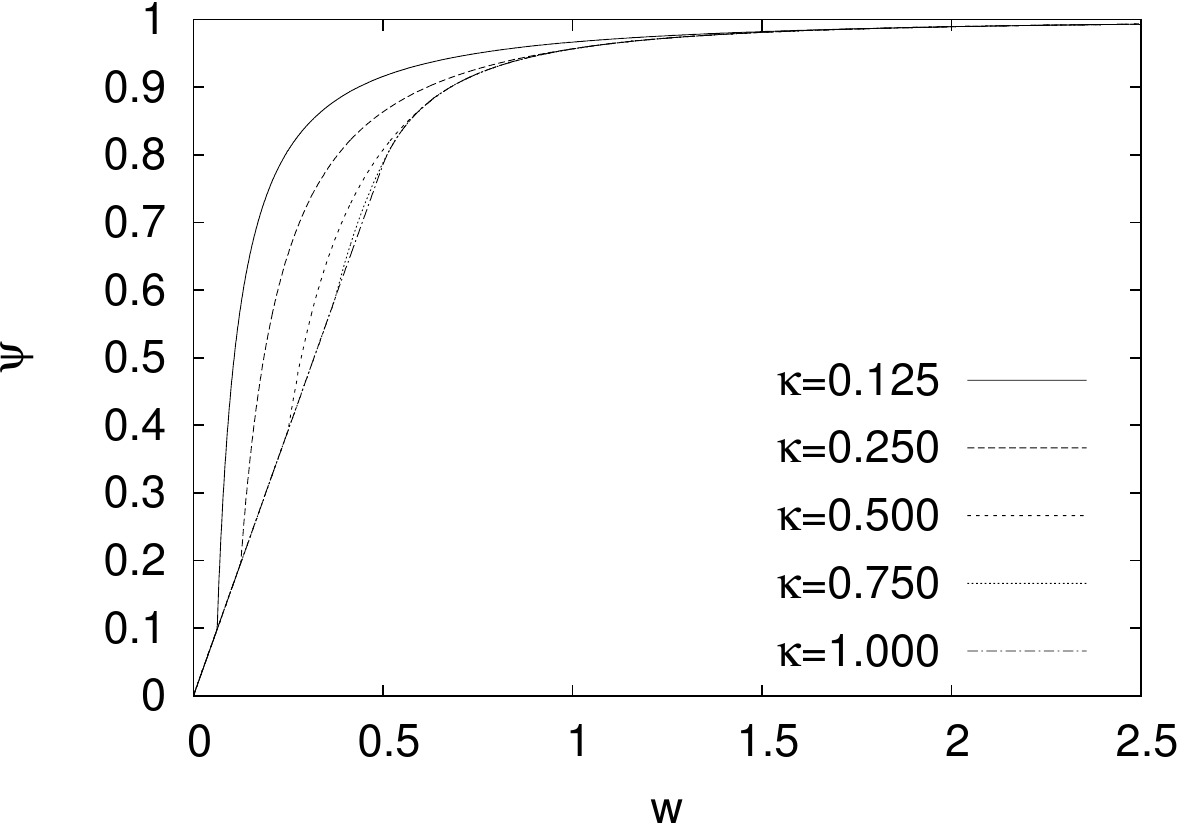}
\caption{On-shell transmission $\psi(w,\kappa)$ through a channels with homogeneous magnetic field as the function of momentum $|p|$ for various $\kappa$.}
\label{pic:mag_trans}
\end{figure}

\begin{figure}[!htb]
\centering
\rotatebox{90}{\footnotesize\hspace{2cm}$\kappa=0.25$}\hskip5mm%
\includegraphics*[width=5.2cm,bb=0 39 230 230]{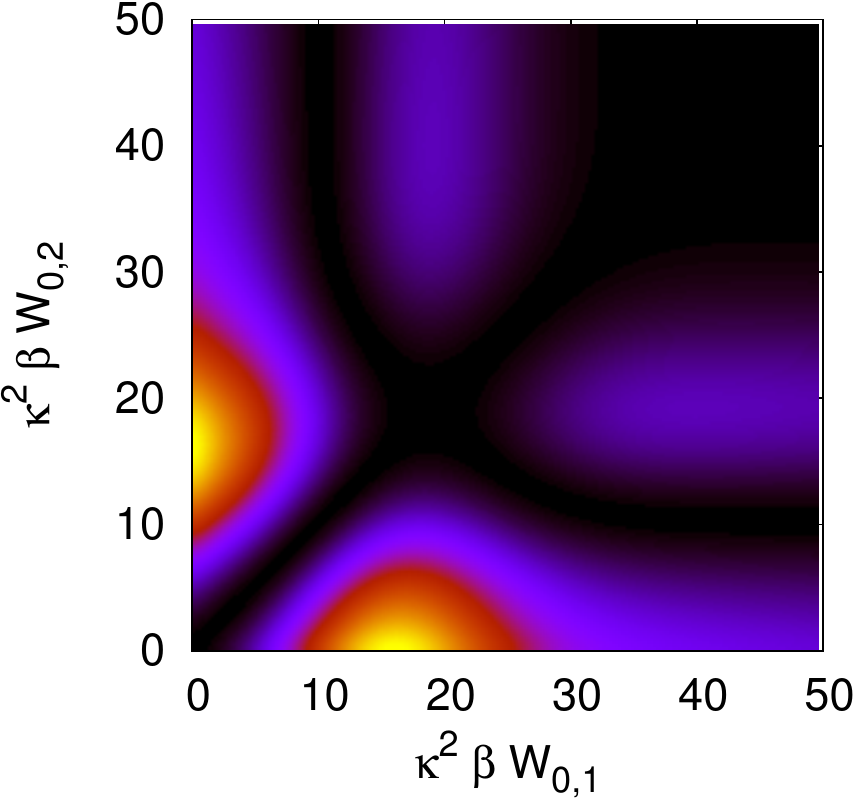}\hskip5pt%
\includegraphics*[width=4cm,bb=53 40 230 230]{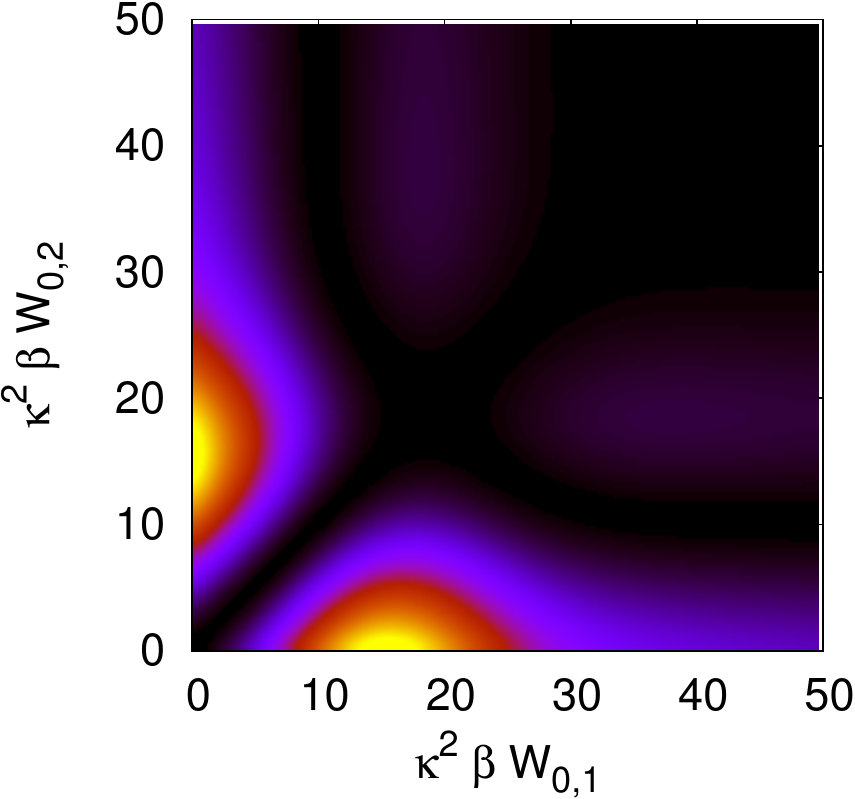}\\
\rotatebox{90}{\footnotesize\hspace{2cm}$\kappa=0.50$}\hskip5mm%
\includegraphics*[width=5.2cm,bb=0 39 230 230]{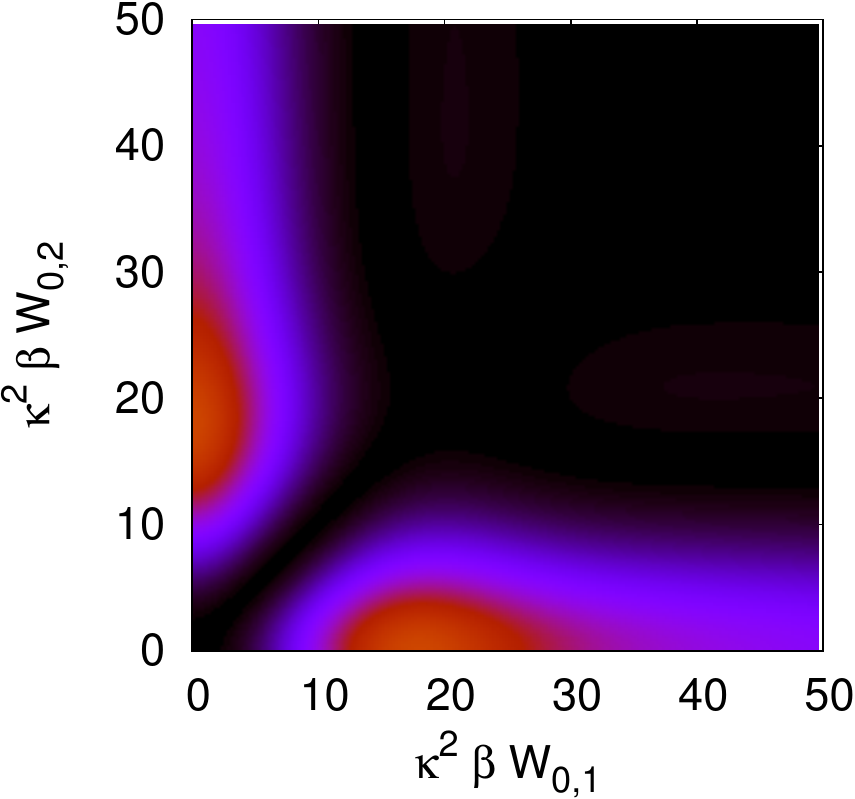}\hskip5pt%
\includegraphics*[width=4cm,bb=53 40 230 230]{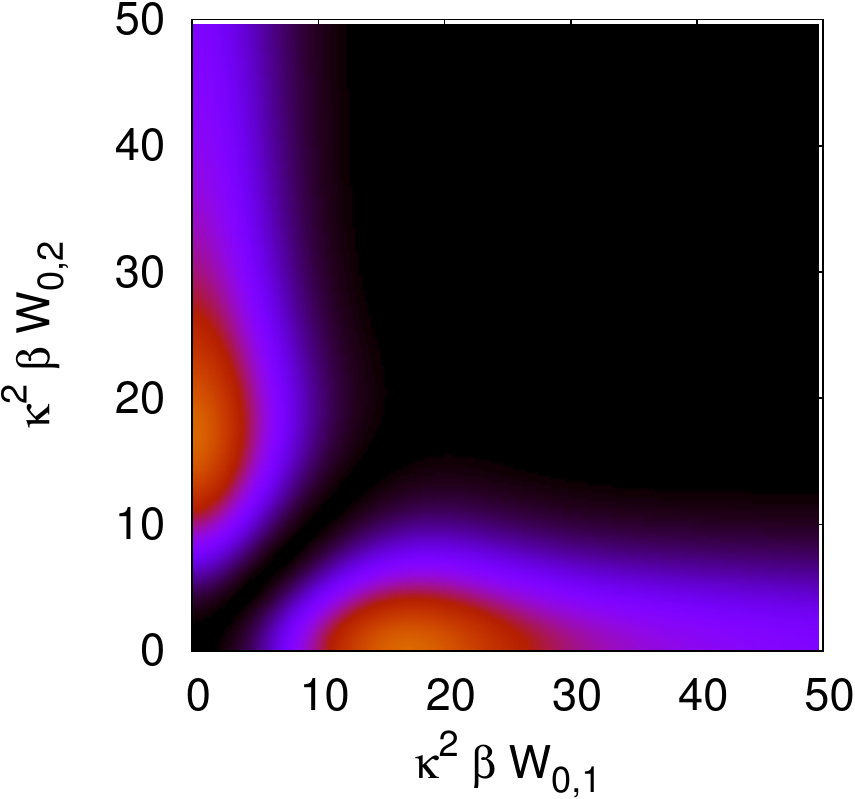}\\
\rotatebox{90}{\footnotesize\hspace{2cm}$\kappa=1.00$}\hskip5mm%
\includegraphics*[width=5.2cm,bb=0 0 230 230]{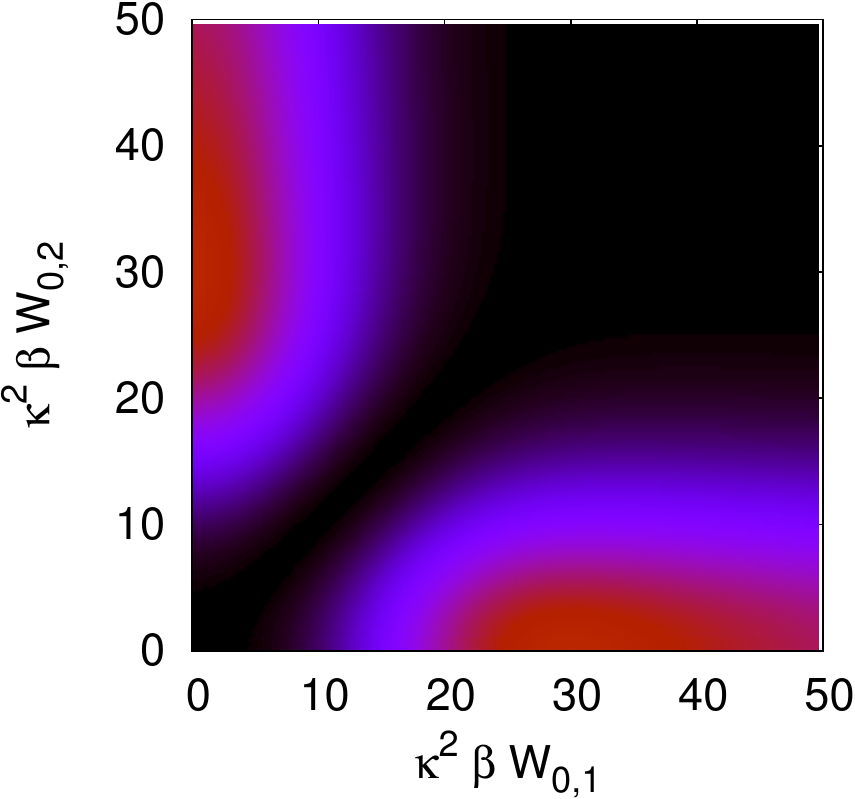}\hskip5pt%
\includegraphics*[width=4cm,bb=53 0 230 230]{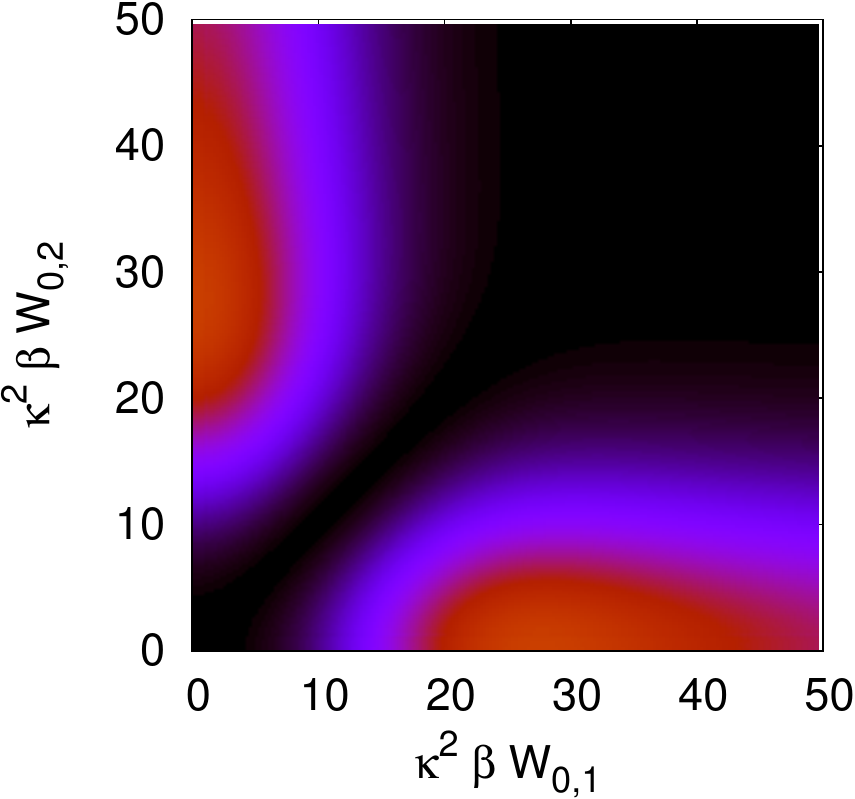}\\[1mm]
\hskip2cm\includegraphics[width=10cm]{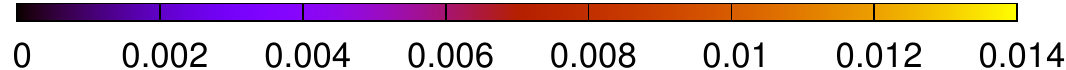}\\
\hspace{2cm}\hbox to12cm{\footnotesize\hfil (a)\hfil\hfil (b) \hfil}
\caption{The relative optimal efficiency $\eta^*/\eta_{\rm Carnot}$ (a) and the corresponding rescaled power $P^*/(\eta_{\rm Carnot}^2 \mean{T})$ as a function of relative energy scales $\beta \kappa^2 W_{0,i}$ at mean temperature $T=1$ for different values of $\kappa$.}
\label{pic:mag_lin}
\end{figure}

\begin{figure}[!htb]
\centering
\includegraphics[width=7.5cm]{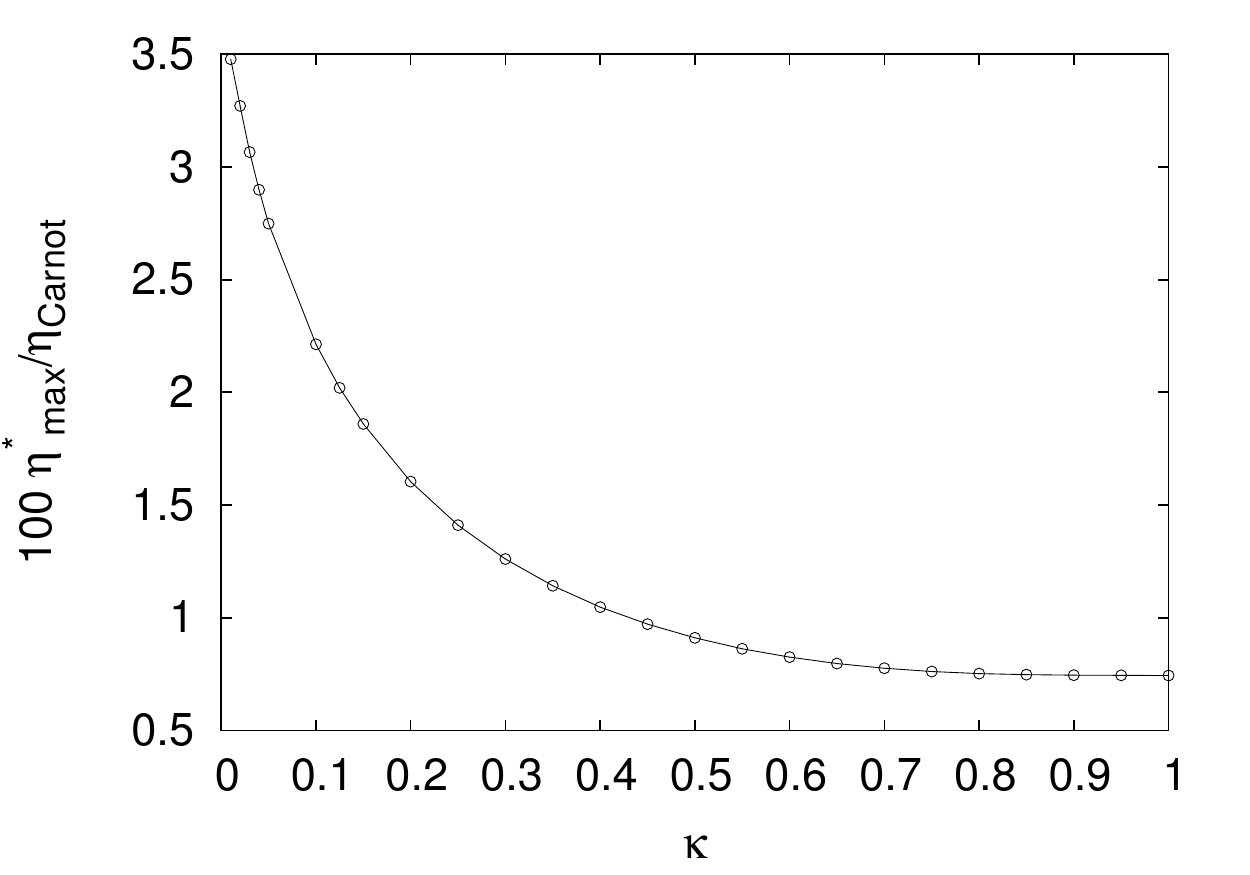}\hskip5pt%
\includegraphics[width=7.5cm]{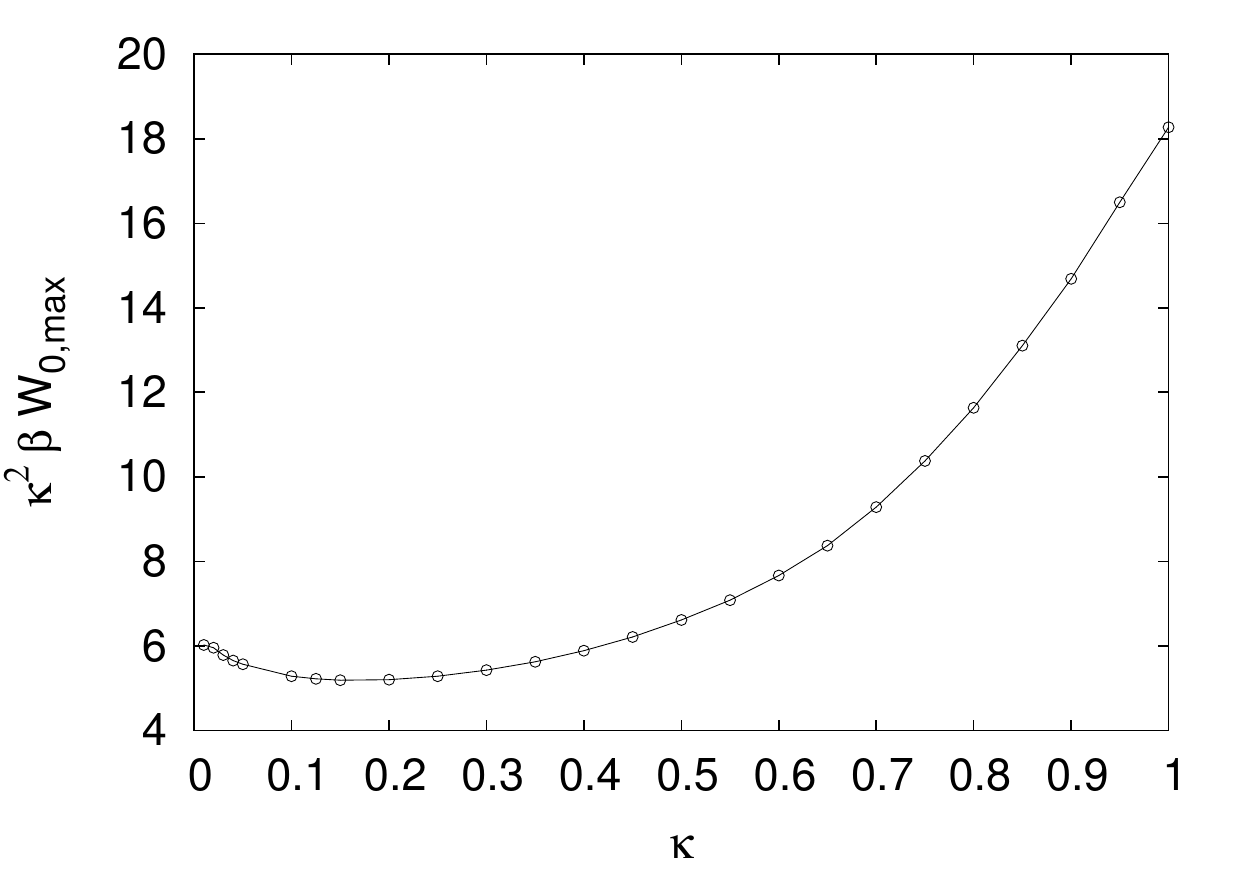}\\
\hbox to14cm{\footnotesize\hfil (a)\hfil (b) \hfil}
\caption{Relative maximal optimal efficiency $\eta^*_{\rm max}/\eta_{\rm Carnot}$ (a) and its position $\beta W_0$ (b) for different values of $\kappa$.}
\label{pic:mag_max}
\end{figure}

In order to discuss the heat engine in the linear temperature difference regime using the linear expansion of currents (\ref{eq:currmat}) we have numerically checked the validity of Eqs. (\ref{eq:lin_cond}). We calculate the transport coefficients $t_i^{(0)}$, $q_i^{(0)}$ and $k_i^{(0)}$ (\ref{eq:coef_gen}) at the mean reciprocal temperature $\beta=1/T$ using the transmission function of the channels at zero electric field $\tau_{\nu,i}|_{U_i=0}$ . The sections without the magnetic and electric field are perfect conductors and therefore transmission through the $i$th channel is determined by the transmission properties of the middle section with the magnetic field of strength $\Omega_i$, length $b_i$ and width $h_i$. Consequently the functions $\tau_{\nu,i}|_{U_i=0}$ in (\ref{eq:coef_gen}) can be replaced by the transmission function of the middle section $\tau_{\rm B}(y, p_x ,p_y; \Omega_i, b_i,h_i)$ and we can write the transport coefficients as 
$$
  (t_i^{(0)}, q_i^{(0)},k_i^{(0)})
  = \frac{1}{h_i} \int_0^{h_i} \dd y \int_0^\infty \dd p_x \int_{-\infty}^{\infty} \dd p_x\,
  P(p; \beta) \tau_{\rm B}(y,p; \Omega_i, b_i,h_i) 
  (1, W(p), W(p)^2) \>.
$$
where $p=(p_x,p_y)$ is the momentum and $W(p) = \frac{1}{2} (p_x^2 + p_y^2)$ is the kinetic energy of the particles  effused into the channels. By closer analysis of dynamics we find that transport coefficients can be conveniently expressed as
\beq 
  (t_i^{(0)}, q_i^{(0)}, k_i^{(0)} )
  = \frac{2 \beta^{\frac{3}{2}}}{\sqrt{\pi}}
  \int_0^\infty \dd W\, W^{\frac{1}{2}} e^{-\beta W} 
  \psi\left( \frac{W}{W_{0,i}}, \kappa_i \right)
  (1,W,W^2)\>,
  \label{eq:mag_mom}
\eeq
with the on-shell transmission probability $\psi$ defined as
$$
  \psi\left(w, \kappa \right)  
  = \frac{1}{2} \int_0^1 \dd y \int_{-1}^1 \dd s\,
  \tau_{\rm B}\left(y,\sqrt{1-s^2}, s, \sqrt{w}, \kappa, 1\right) \>,
$$
where we introduce the energy scale 
$
 W_{0,i} = \frac{1}{2} (\Omega_i h_i)^2
$
and the ratio between the length of the section and its width 
$
  \kappa_i = b_i/h_i\>.
$ 
The energy scale is proportional to the ratio between the critical temperature as defined in the paper \cite{casati07} and the average bath temperature $T$. In general $\psi(w,\kappa)$ is a piecewise smooth function of both parameters as we can see in figure (\ref{pic:mag_trans}). Shorter is the section in comparison to the width, higher is the average transmission probability.
As expected, the transmission probability at given $w$ increases with decreasing $\kappa$. If the section with the magnetic field is longer than its width, then $\psi$ becomes independent of $\kappa$, meaning
$$
  \psi(w,\kappa) = \psi(w, 1) \qquad {\rm for}\quad \kappa>1\>.
$$
Additionally, $\psi(w,\kappa)$ as a function $w$ is independent of $\kappa$ in the limit of small and large $w$. The asymptotic expansion in these two limits reads
$$ 
  \hspace{-10mm}\psi(w, \kappa)\sim \frac{\pi}{2} w^{\frac{1}{2}}  \quad   w\to 0 
  \quad{\rm and}\quad
  \psi(w, \kappa) \sim 1 - \frac{1}{24}w^{-1} + O(w^{-2}) \quad  w\to \infty\>.
  \label{eq:psi_asymp}
$$ 
In the absence of an electric field, $\psi$ can be obtained analytically, but it is difficult. Nevertheless, we work here with a numerically computed $\psi$ enhanced by analytical approximations. 
By using $\psi$ to calculate coefficients (\ref{eq:mag_mom}) we study the optimal efficiency $\eta^*$ and the corresponding power $P^*$ of the heat engine given by equations (\ref{eq:eta_opt}) and  (\ref{eq:power_opt}), respectively, for different value of ratio $\kappa$. 
The results are shown in figure \ref{pic:mag_lin}, where we plot $\eta^*$ and $P^*$ as a function of the rescaled energy scales $\kappa^2\beta W_{0,i}$.
Because the channels are identical, $\eta^*$ and $P^*$ are symmetrical on exchange of parameters. The areas of high optimal efficiency $\eta^*$ and high power $P^*$ overlap and are located along the abscissa and ordinate axes in $W_{0,1}\times W_{0,2}$ parameter space. For a given $\kappa$, the maximum of the optimal efficiency is achieved on lines $(W_{0,1},0)$ and $(0,W_{0,2})$. The position of maximal optimal efficiency $W_{0,\rm max}$ and its value $\eta^*_{\rm max}$ depend on $\kappa$, as we see in figures \ref{pic:mag_max}.a and \ref{pic:mag_max}.b.
The maximal efficiency $\eta^*_{\rm max}$ and the relative energy scale $\beta W_{\rm 0,max}$ defining its position are increasing as we decrease $\kappa$.  Numerical calculations of $\eta^*_{\rm max}$ are difficult to perform for $\kappa$ near to zero. In order to obtain results in this limit, we approximate the on-shell transmission probability with its approximate 
limiting form, given by
$$
  \psi(w,\kappa)  \approx  \psi_a(w,\kappa) = \left \{
  \begin{array}{lll}
  \frac{\pi}{2} w^{\frac{1}{2}} &:& w \le \frac{1}{\pi^2}(1- \sqrt{1-\pi \kappa})^2 \\
  1 - \frac{1}{2}\kappa w^{-\frac{1}{2}} &:& {\rm otherwise}
  \end{array} \right.\>,
$$
within numerically determined absolute error $|\psi(w,\kappa)-\psi_a(w,\kappa)| < 5\cdot 10^{-3}$ for $\kappa < 10^{-2}$. By using this approximation we see that the maximal optimal efficiency increases with decreasing $\kappa$ and converges in the limit $\kappa\to 0$ to its upper bound
$$
  \left. \frac{\eta_{\rm max}^*}{\eta_{\rm Carnot}}\right|_{\kappa=0}
  = 0.0373 \pm 0.001
$$
at which the relative energy scale is $\kappa^2\beta W_0 \sim 6.55 \pm 0.01$. Numerical results in figure \ref{pic:mag_max}.b support this finding. We see that the maximal possible efficiency in the linear regime is below 4\% of that achieved in the Carnot's cycle.

\subsection{Non-linear temperature difference regime}
In the non-linear temperature regime, where $|T_\rL-T_\rR| \gtrsim \mean{T}$, we numerically obtained the transport coefficients $t_{\nu,i}$ and $q_{\nu,i}$ (\ref{eq:gen_tranp_coef}) and study the performance of the heat engine. In order to obtain heat engine's optimal performance at fixed setup, we numerically find the  optimal efficiency w.r.t. the electric potential $U_i$. \par
\begin{figure}[!htb]
\centering
\includegraphics[width=6cm]{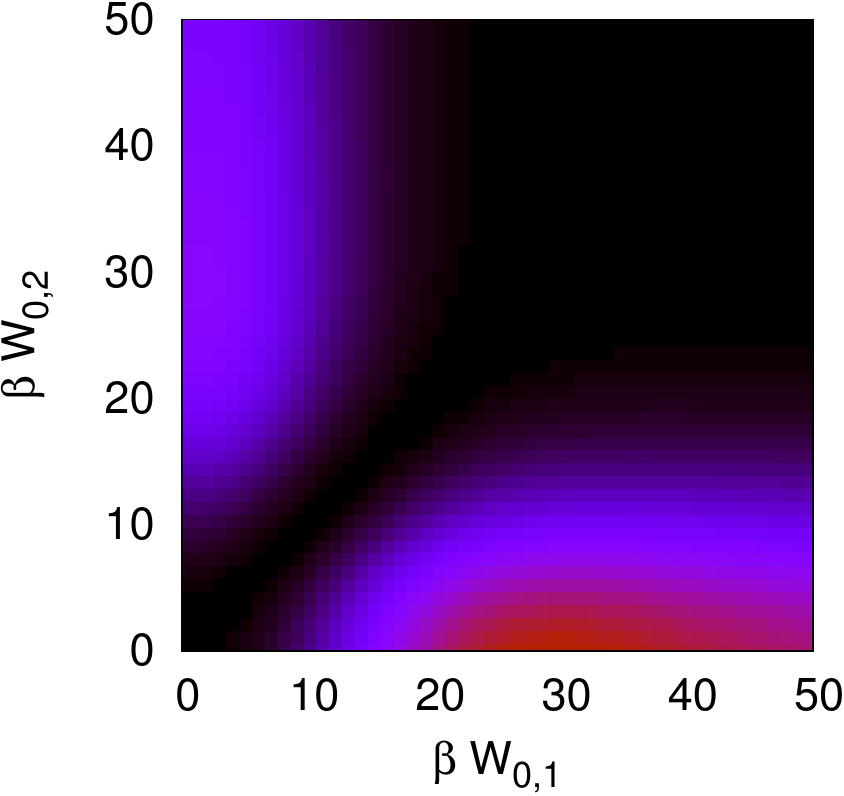}\hskip5mm%
\includegraphics[width=6cm]{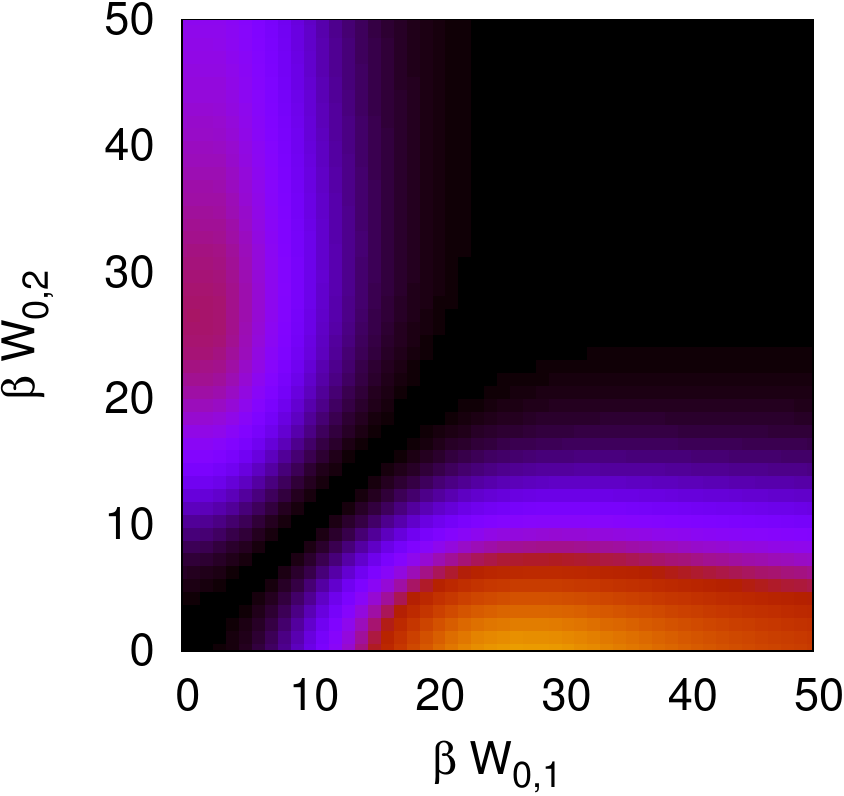}\\[1mm]
\hskip15mm\includegraphics[width=10cm]{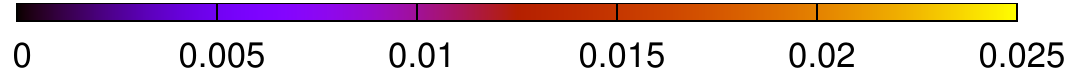}\\
\hskip2cm\hbox to12cm{\footnotesize\hfil (a)\hfil\hfil (b) \hfil}
\caption{The relative optimal efficiency $\eta^*/\eta_{\rm Carnot}$ (a) and the corresponding rescaled power $P^*/(\eta_{\rm Carnot}^2 \mean{T})$ as a function of relative energy scales $\beta W_{0,i}$ for temperatures $T_\rL=0.55$ and $T_\rR=1.45$ and their mean $\mean{T}=1$.}
\label{pic:mag_nonlin}
\end{figure}

As an example, we present in figure \ref{pic:mag_nonlin} the optimal efficiency $\eta^*$ and corresponding power $P^*$ as functions of relative energy scales $\beta W_{0,i}$ for $i=1,2$. In the example the average temperature is $\mean{T}=1$, the electric field is applied only in the first channel and the lengths of different sections and widths of the channel are equal $a_i=b_i=c_i= h_i =1$. In this setup we fix $\kappa_i = 1$. 
The $\eta^*$ and $P^*$ as the functions of $\beta W_{0,i}$ are slightly asymmetric, due to application of bias voltage to the first channel only. Their functional dependence is not much different than that obtained in the linear temperature regime. We find that the relative maximal optimal efficiency $\eta^*_{\rm max}/\eta_{\rm Carnot}$ slightly increases when we increase the temperature difference between the baths. By increasing the temperature difference, at fixed average temperature, we also increase the power and heat current since they are approximately proportional to $\eta_{\rm Carnot}$ and $\eta_{\rm Carnot}^2$, respectively.
 
\section{Conclusions}

We have demonstrated the operation of a thermoelectric heat engine and refrigerator in terms of a classical mechanical model, which combines the deterministic classical scattering dynamics and stochastic heat reservoirs. The advantage of our simple model is that it is analytically treatable and that it allows for explicit exact results. In particular we have been able to disclose various situations in which the efficiency can be close to Carnot's and thus pose a promising alternative for technological application, in particular at a nanoscale, where the motion between the baths could be approximated by dissipationless dynamics.

\section*{Acknowledgements}

We acknowledge financial support by the research grants  Z1-0875 (MH) and J1-2208, P1-0044 (TP) of the Slovenian Research Agency (ARRS), and
MIUR-PRIN 2008 and by Regione Lombardia The authors thank the Max Planck Institute for the Physics of Complex Systems in Dresden for the hospitality in Advanced study group 2010 during the finalization of the paper.

\appendix

\section{Calculating the classical scattering matrix}

The channels joining thermo-chemical baths are treated as two-port scatterers for particles of $d$ degrees of freedom. Let us consider the $i$th channel connected to the left and to the right bath through openings in space $\Gamma_{\rL, i}$ and $\Gamma_{\rR, i}$, respectively, of an equal area $A_{\rL, i} = A_{\rR, i} =: A_i$. The transport properties of the channels are controlled via homogeneous electric field with the voltage $U_i$ between the ends. We image having a stationary transport in the channel with particles being injected and ejected at both sides at constant rate. The probability density of particle's positions and momenta that are injected at $\nu$ side is given by 
$$
  \rho_{\rm in, \nu}(x,p)\quad (x,p)\in\Gamma_{\nu, i}\times M_\nu\>,
$$
where we introduce a momentum space for particles traveling in $\nu$ direction $M_\nu = \bR_{\sigma_\nu} \times \bR^{d-1}$. Similarly, the probability density of particles ejected at side $\nu$ is 
$$
 \rho_{\rm out, \nu}(x,p)\quad (x,p)\in\Gamma_{\nu, i}\times M_{\neg\nu}\>.
$$
We assume $\rho_{\rm in, \nu}(x,p)=0$ for $\sigma_\nu p_1< 0 $ and $\rho_{\rm out, \nu}(x,p)=0$ for $\sigma_\nu p_1 >0 $. Using the introduced notation we may write the probability density of ingoing particles $\rho_{\rm in}$ on the phase space  $V_{\rm in,i} = (\Gamma_{\rL, i}\times M_\rL) \cup (\Gamma_{\rR, i}\times M_\rR)$ and the probability density of outgoing particles $\rho_{\rm out}$ on the phase space $V_{\rm out,i} = (\Gamma_{\rL, i}\times M_\rR) \cup (\Gamma_{\rR, i}\times M_\rL)$ are defined as
$$
  \rho_{\rm in}= \rho_{\rm in, \rL} + \rho_{\rm in, \rR}
  \quad {\rm and} \quad
  \rho_{\rm out}= \rho_{\rm out, \rL} + \rho_{\rm out, \rR}\>,
$$
respectively. The two-port classical scattering operator $\hat S$ associated to the channel is a linear map of $\rho_{\rm in}$ into $\rho_{\rm out}$:
$$
  \rho_{\rm out} = \hat S \rho_{\rm in}\>.
$$
The operator $\hat S$ depend on dynamical properties of the channel. By using projectors $\hat P_\nu$ defined as $P_\nu f = \theta(\sigma_\nu p_1) f$, we can decompose the operator $\hat S$ into four operators:
$$
  \hat T_\rL  = \hat P_\rL \hat S \hat P_\rL \>, \quad 
  \hat R_\rL  = \hat P_\rR \hat S \hat P_\rL \>, \quad
  \hat T_\rR  = \hat P_\rR \hat S \hat P_\rR \>, \quad
  \hat R_\rR  = \hat P_\rL \hat S \hat P_\rR \>.
$$
These operators describe (listed in the same order as above from left to right)  the transmission from the left to the right side, the reflection from the left back to the left side, the transmission from the right to the left side and the reflection from the right back to the right side. Notice the sum of these operators is again the operator $\hat S$. The dynamics is deterministic and pointwise. Consequently, the trajectory's entry point $z = (x,p)$ can be uniquely connected to the exit point $z'=(x',p')$ via a map $z'=\phi(z)$ and $\hat S$ is a Koopman operator. This means $\hat S \delta_z' = \delta_{\phi(z')}$, where we use Dirac delta $\delta_z'(z) := \delta(z-z')$. By using  these facts we express the transmission function $\tau_{\nu,i}$ (\ref{eq:trans_def}) as
\beq
  \tau_{\nu,i} (x', p', U_i) 
  = \int_{V_{\rm out, i}} \dd z\, (\hat T_\nu \delta_{z'})(z) 
  = \int_{\Gamma_{\nu, i}\times M_{\neg\nu} } \dd z\, (\hat S \delta_{z'})(z) \>.
  \label{eq:tau_S}
\eeq
The transmission function $\tau_{\nu,i}$ of a channel is an essential ingredient of the presented theory of heat engines and refrigerators, and relation (\ref{eq:tau_S}) connects it with a scattering operator $\hat S$. As we will show in the following, the operator $\hat S$ can be systematically obtained for structurally complicated channels, when scattering in its parts is known.\par
Let us assume that a channel can be divided into two parts with a common cross-section and simpler dynamics, which can be described by the scattering operators $\hat S_1$ and $\hat S_2$. The operator $\hat S_\nu$ can be decomposed into transmission and reflection operators $\hat T_{i,\nu}$ and $\hat R_{i,\nu}$. By knowing them, we can construct the transmission and reflection operators $\hat T_{\nu}$ and $\hat R_{\nu}$ corresponding to the scattering operator $\hat S$ of the whole channel via the concatenation formulas
\begin{eqnarray*}
  \hat R_\rL =  \hat R_{1, \rL} + 
    \hat T_{1, \rR} \hat R_{2,\rL} {\hat L}^{-1} \hat T_{1,\rL}\>, &
  \hat T_\rL = \hat T_{2,\rL} \hat L^{-1} \hat T_{1,\rL}\>, \\
  \hat R_\rR = \hat R_{2,\rR} + 
    \hat T_{2,\rL} \hat R_{1,\rR} {\hat {L'}}^{-1} \hat T_{2,\rR}\>, &
  \hat T_\rR = \hat T_{1,\rR} \hat {L'}^{-1} \hat T_{2,\rR} \>, 
\end{eqnarray*}
with $\hat L = 1 - \hat R_{1,\rR} \hat R_{2,\rL}$ and $\hat L'=1 - \hat R_{2,\rL} \hat R_{1,\rR}$. This is analogous to concatenation of quantum scattering matrices discussed in \cite{horvat08}. Note that the inverse operator $(1-\hat A)^{-1}$ is just an symbolic abbreviation for the series $\sum_{i=0}^\infty \hat A^i$. In order to simplify the writing we represent the concatenation of two scattering operators $\hat S_1$ and $\hat S_2$ into a resulting scattering operator $S$ by a non-linear and non-commutative product labeled by the symbol $\odot$ yielding
$$
  \hat S = \hat S_1 \odot \hat S_2 \>.
$$
For example, let us consider a straight empty one dimensional channel of length $L$ in the middle of which is a point scatterer that maps the incoming particle momentum $p$ into $p'= O(\epsilon) p$, where $O$ orthogonal matrix depending on the energy $\epsilon = \frac{1}{2} p^T p$. Such maps obeys the energy conservation. The scattering operator associated with the empty wires of length $\frac{1}{2}L$ with the voltage $\frac{1}{2}U$ between ends is defined as
$$
  (\hat E \phi)(p_1,p_2,\ldots,p_d)= 
  \theta(-a)\phi(-p_1,p_2,\ldots,p_d) +
  \frac{|p_1|\theta(a)}{\sqrt{a}} \phi(\sign(p_1)\sqrt{a},p_2,\ldots,p_d)
$$
with $a = p_1^2 + \sign(p_1) U$. The scattering operator corresponding to the scatterer in the middle of the channel is written as
$$
  (\hat P \phi)(p) = \phi(O(\epsilon)^{-1}p)\>.
$$
Then we can write the scattering operator corresponding to the whole channel as a product
$$
  \hat S = \hat E \odot \hat P \odot \hat E \>.
$$
The latter can be calculated systematically. Such kind of channels are used for conduction of one-dimensional particles, where $O(\epsilon) \in \{-1,1\}$, in a realization of a heat engine discussed in section \ref{sec:1dwindow}.

\section{Relaxation times}

We consider a heat engine or refrigerator with multiple channels and no particles initially in the channels. Then at time $t=0$, the particles start to be injected into the channels from the left and right bath with the injection rates $\gamma_\rL$ and $\gamma_\rR$, respectively. The number of particles in channels $N(t)$ increases with time. Eventually it saturates to a stationary value and so the heat engine enters into a non-equilibrium stationary state. The characteristic time scales of this relaxation process are essential for practical implementations of heat engines and refrigerators. These can be obtained through the study of functions $N(t)$, defined more precisely in the following.\par
Let us introduce time function $\phi_{\nu,i}$, where $\phi_{\nu,i}(q,p)\ge 0$ represents the time needed for a particle entering $i$th channel from side $\nu$ at the point $(q,p) \in {\cal V}_{\nu,i}=\Gamma_{\nu,i}\times M_\nu$ to return into the baths. The time function can be measured for channels and in some simple cases even analytically expressed. The number of particles in $i$th channel coming form side $\nu$ is given by
\beq
\hspace{-5mm}
  N_{\nu,i}(t) = 
  \frac{1}{A_{\nu, i}}\int_{\Gamma_{\nu,i}} \dd q \int_{\bR^d} \dd p\,
   P(p,\sigma_\nu,\beta_\nu)
  \tau_{\nu,i}(q,p,U_i) \min(\phi_{\nu,i}(q,p), t)\>.
  \label{eq:nr_t}
\eeq
Then the total number of particles in the channels is equal to
$$
  N(t) = \sum_{\nu,i} r_i\gamma_\nu N_{\nu,i}(t)\>.
$$
In order to obtain the relaxation time scales we need to understand asymptotics of $N(t)$, which is determined by effective injection rates $r_i\gamma_\nu$ and asymptotics of individual contributions $N_{\nu,i}$. It is difficult to say something definite about the latter. But if $\phi_{\nu,i}$ is continuous except on ${\cal U}_{\nu,i} \subset {\cal V}_{\nu,i} $ of zero-measure then $N_{\nu,i}$ is also continuous. From definition (\ref{eq:nr_t}) it is clear, that $N(t)$ is non-constant in the limit $t\to\infty$, if $\phi_{\nu,i}$ has a singularity on subset of ${\cal U}_{\nu,i}$. This nontrivial asymptotics of $N(t)$ is determined by the functional dependence of $\phi_{\nu,i}$ in the vicinity of ${\cal U}_{\nu,i}$.\par
As an illustration, let us consider an empty one-dimensional channel of length $L$ with one-dimensional particles injected from one side at the rate $\gamma$. The time function $\phi$ of such channel depends on the voltage $U$ and reads
$$
  \phi(p) = \left \{
  \begin{array}{lll}
  \frac{2 L}{|p| + \sqrt{p^2 - u}} &:& u < 0 \\
  \theta(\sqrt{u} - |p|)\frac{2|p| L}{u} + 
  \theta(|p| - \sqrt{u}) \frac{2 L}{p + \sqrt{p^2 + u}}
  &:& {\rm otherwise} 
  \end{array}  
  \right. \>,
$$
with $u = 2\sign(p)U$. By using $\phi$ we can compute the number of particles in the channel coming form one side $N_1(t)$ as a function of time $t$. In the presence of the electric field $\phi$ is bounded from above by $t_0 = L/\sqrt{2|U|}$, which represents the time needed to achieve the stationary state. The stationary number of particles is $N_0:=\lim_{t\to\infty} N_1(t) \propto \gamma t_0$. In the absence of the field $\phi(p)$ does not have an upper bound and behaves as $\phi(p) \asymp p^{-1}$ near $p=0$. Consequently, the number of particles $N_1$ converges slowly in algebraic manner as
$$
  N_1(t) \asymp 
  N_0 + 
  \frac{t_1}{t} \left(\gamma t_1 - \frac{2 N_0}{\sqrt{\pi}}\right)+ 
  O(t^{-3})\>, 
$$
with the stationary number $N_0 = \gamma L \sqrt{\beta \pi/2}$ and the time scale of relaxation $t_1 = L \sqrt{\beta/2}$. In view of these results, we expect in general for channels composed of parallel walls that in the absence of the field the convergence towards the stationary state is algebraic. Therefore to achieve the stationary state in a finite time in such cases it is practical to apply electric fields in all channels.

\end{document}